\documentclass[12pt]{article}
\usepackage[super,comma,numbers,square]{natbib}
\usepackage{graphicx}
\usepackage{float}
\usepackage[caption=false]{subfig}
\usepackage[a4paper,margin=1in]{geometry}
\usepackage{times}
\usepackage{amsmath}
\usepackage{xspace}
\usepackage{amssymb}
\usepackage{xr}
\usepackage[]{hyperref}

\newcommand{\spacing}[1]{\renewcommand{\baselinestretch}{#1}\large\normalsize}
\spacing{1.25}

\newenvironment{affiliations}{%
	\setcounter{enumi}{1}%
	\setlength{\parindent}{0in}%
	\slshape\sloppy%
	\begin{list}{\upshape$^{\arabic{enumi}}$}{%
			\usecounter{enumi}%
			\setlength{\leftmargin}{0in}%
			\setlength{\topsep}{0in}%
			\setlength{\labelsep}{0in}%
			\setlength{\labelwidth}{0in}%
			\setlength{\listparindent}{0in}%
			\setlength{\itemsep}{0ex}%
			\setlength{\parsep}{0in}%
		}
	}{\end{list}\par\vspace{12pt}}

\renewenvironment{abstract}{%
	\setlength{\parindent}{0in}%
	\setlength{\parskip}{0in}%
	\bfseries%
}{}%{\par\vspace{-6pt}}

\newenvironment{addendum}{%
	\setlength{\parindent}{0in}%
	\small%
	\begin{list}{Acknowledgements}{%
			\setlength{\leftmargin}{0in}%
			\setlength{\listparindent}{0in}%
			\setlength{\labelsep}{0em}%
			\setlength{\labelwidth}{0in}%
			\setlength{\itemsep}{12pt}%
			}
	}
	{\end{list}\normalsize}

\newcommand{\smv}{Supplementary Video}
\newcommand{\sfig}{Supplementary Fig.}
\newcommand{\efig}{Extended Data Fig.}
\newcommand{\efigs}{Extended Data Figs.}
\newcommand{\snote}{Supplementary Note\xspace}
\newcommand{\ion}[2]{#1$\;${\small\rmfamily\rom{#2}}\relax}%
\newcommand*{\rom}[1]{\uppercase\expandafter{\romannumeral #1\relax}}

\newcommand{\aap}{    {\it Astron. Astrophys.}}

\newcommand{\aapr}{   {\it Astron. Astrophys. Rev.}}

\newcommand{\apj}{    {\it Astrophys. J.}}
\newcommand{\apjl}{   {\it Astrophys. J. Lett.}}

\newcommand{\jgr}{    {\it J. Geophys. Res.}}

\newcommand{\solphys}{{\it Solar Phys.}}
\newcommand{\ssr}{    {\it Space Sci. Rev.}}

\begin{document}

\title{Complete replacement of magnetic flux in a flux rope \\ during a coronal mass ejection} 

\author{Tingyu Gou$^{1*}$, Rui Liu$^{1,2*}$, Astrid M. Veronig$^{3}$, Bin Zhuang$^{4}$,\\
	    Ting Li$^{5,6}$, Wensi Wang$^{1,7}$, Mengjiao Xu$^{1,8}$, Yuming Wang$^{1,2,8}$ }

\date{\vspace{-5ex}}

\maketitle

\hspace{2pt}

\begin{affiliations}
	%\small
	\item School of Earth and Space Sciences/CAS Center for Excellence in Comparative Planetology/CAS Key Laboratory of Geospace Environment, University of Science and Technology of China, Hefei 230026, China
	\item Collaborative Innovation Center of Astronautical Science and Technology, Hefei, China
	\item Institute of Physics \& Kanzelh\"{o}he Observatory for Solar and Environmental Research, University of Graz, 8010 Graz, Austria
	\item Institute for the Study of Earth, Oceans, and Space, University of New Hampshire, Durham, NH 03824, USA
	\item CAS Key Laboratory of Solar Activity, National Astronomical Observatories, Beijing 100101, China 
	\item School of Astronomy and Space Science, University of Chinese Academy of Sciences, Beijing 100049, China
	\item Mengcheng National Geophysical Observatory, University of Science and Technology of China, Mengcheng 233500, China
	\item Deep Space Exploration Laboratory, Hefei 230026, China \\
	*E-mail: tygou@ustc.edu.cn; rliu@ustc.edu.cn
\end{affiliations}

\begin{abstract}
Solar coronal mass ejections are the most energetic events in the Solar System. In their standard formation model, a magnetic flux rope builds up into a coronal mass ejection through magnetic reconnection that continually converts overlying, untwisted magnetic flux into twisted flux enveloping the pre-existing rope. However, only a minority of coronal mass ejections carry a coherent magnetic flux rope as their core structure, which casts doubt on the universality of this orderly wrapping process. Here we provide observational evidence of a different formation and eruption mechanism of a magnetic flux rope from an S-shaped thread, where its magnetic flux is fully replaced via flare reconnections. One of the footpoints of the sigmoidal feature slipped and expanded during the formation, and then moved to a completely new place, associated with the highly dynamical evolution of flare ribbons and a twofold increase in magnetic flux through the footpoint, during the eruption. Such a configuration is not predicted by standard formation models or numerical simulations and highlights the three-dimensional nature of magnetic reconnections between the flux rope and the surrounding magnetic field.
\end{abstract}

\section*{Introduction}

Magnetic flux ropes (MFRs), which consist of twisted magnetic field lines, play a key role in understanding various phenomena in astrophysical, space and laboratory plasmas. In solar physics, MFRs have been widely accepted as a fundamental structure in solar eruptions \cite{Patsourakos2020,LiuR2020}, which are accompanied by a rapid release of a huge amount of energy  \cite{Priest&Forbes2002}. Driven by solar eruptions, the space weather has become increasingly important for various technological systems in the modern society \cite{Temmer2021}, which makes it essential to study the genesis and eruptive mechanism of solar MFRs.

MFRs on the Sun manifest as various observational features, such as filaments, hot channels, coronal cavities, and S-shaped loop bundles known as sigmoids \cite{ChengX2017,Schmieder2015,Green2018}. A coherent MFR could either be present prior to the eruption \cite{ZhangJ2012,Patsourakos2013} or form by magnetic reconnection within a sheared arcade during the eruption \cite{Moore2001,Karpen2012}. The two different scenarios, which refer to two competing pre-eruption magnetic field configurations of solar eruptions, i.e., MFR versus sheared magnetic arcade \cite{Patsourakos2020}, have been attracting great interest and under intense debate for decades. Recent observations reveal that a coherent MFR may build up upon a core `seed' that is initiated from the coalescence of mini plasmoids in a pre-existing current sheet\cite{Gou2019}, which gives a specific intermediate state between the two distinct configurations,  i.e., a `hybrid' state with a twisted core embedded in a sheared arcade \cite{LiuR2020}.

The standard flare model describes the eruption of a pre-existing MFR, in which magnetic reconnection in the current sheet underneath adds field lines wrapping around the rising rope \cite{Kopp&Pneuman1976,Lin&Forbes2000}, producing post-flare loops connecting two parallel flare ribbons on the surface. In consideration of three-dimensional (3D) reconnection occurring at quasi-separatrix layers (QSLs) that separate the MFR from the surrounding untwisted fields\cite{Demoulin2006}, the two-dimensional (2D) standard model is extended to three dimensions, with the (quasi-)separatrix footprints of coronal magnetic fields mapped by double J-shaped flare ribbons \cite{Aulanier2012,Aulanier2013,Janvier2013}: the straight sections of flare ribbons correspond to the footprints of the hyperbolic flux tube (a collection of two intersecting QSLs\cite{Titov2002}) below the MFR, which is the 3D counterpart of the vertical current sheet in the 2D model; the endmost hooks correspond to the footprints of the QSLs wrapping around the MFR, therefore outlining its two feet anchored in the dense photosphere. Inside the hooks, coronal emission dims due to the evacuation of coronal matter along the rope's legs during the eruption \cite{Janvier2014,WangWS2017,Veronig2019,XingC2020}. Thus, as the chromospheric imprint of coronal reconnections, motions and morphological changes of flare ribbons reflect the changes of coronal magnetic connectivities and the evolution of eruptive MFRs.

After solar eruptions, MFRs are sometimes detected in situ in interplanetary space and termed magnetic clouds (MCs), which are characterized by a large and smooth rotation in the field's direction, enhanced magnetic field strength, a depressed proton temperature and density, and a low plasma $\beta$ \cite{Burlaga1981,Klein&Burlaga1982}. However, only about one third of interplanetary coronal mass ejections (ICMEs) possess typical MFR features \cite{ChiYT2016}, in spite of their same origin from the Sun. This mismatch between the in-situ observation and the anticipation that all CMEs are essentially MFRs \cite{Vourlidas2013} is traditionally attributed to interactions among the ejecta and the solar wind as well as to unfavorable spacecraft trajectories within ejecta \cite{SongHQ2020}. But are there also other physical mechanisms that can account for some of the missing of well-structured MFRs that manifest as complex ejecta in interplanetary space?

Here we study the formation and eruption of a solar MFR in a sigmoidal active region. The combination of the low-atmosphere and coronal observations highlights the buildup of an MFR before the eruption and its subsequent restructuring during the eruption. The lower-atmosphere counterpart of the erupting structure exhibits a highly dynamic evolution while one of its feet migrates to a completely new place compared to the pre-eruption MFR, which indicates a complete replacement of magnetic fluxes in the rope. The restructuring of the MFR highlights the importance of 3D reconnections with the surrounding coronal magnetic fields during its the development into a CME, and sheds light on the implication of the complex dynamics occurring in the solar source active region for interplanetary ejecta.

\section*{Results}

\subsection*{Overview of the event}

The eruption of interest occurs in the NOAA active region 12158 on 2014 September 10 near the solar disk center. The active region consists of a leading positive-polarity sunspot surrounded by negative polarities in the southeast, and it exhibits a typical sigmoidal configuration (Fig.~\ref{fig:ov}; \smv~1). A bundle of inverse-S shaped loops is observed several hours before the eruption in the SDO/AIA (Methods) 94~\AA\ channel (Fig.~\ref{fig:ov}d; \smv~2). The eruption starts with the rise of a coherent sigmoidal structure as observed in AIA's hot passbands (e.g., 131~\AA; Fig.~\ref{fig:ov}b; \smv~1), which is interpreted as an MFR in previous studies on the event \cite{ChengX2015,Dudik2016,ZhaoJ2016}. It produces a GOES X1.6-class (Methods) flare starting at 17:21~UT, and a fast halo coronal mass ejection (CME), which is detected as an ICME near the Earth two days after (\snote~1.3). 

A sigmoidal MFR builds up from a thin thread in the core of a sheared magnetic arcade during the flare precursor phase (Fig.~\ref{fig:ov}a,b), accompanied by footpoint brightening in the low solar atmosphere. The western foot of the pre-eruption MFR is located in positive polarities near the sunspot (Fig.~\ref{fig:ov}g); the eastern foot in negative polarities is outlined by a low-atmosphere ribbon-like brightening (Fig.~\ref{fig:ov}b,e). During the eruption, the same ribbon undergoes rapid and complex changes (Fig.~\ref{fig:ov}c,f--h). Below we study in detail the formation of the sigmoid and its complex evolution during the eruption.

\subsection*{Appearance of the seed MFR}

The eruptive sigmoid originally appears as a brightening inverse-S shaped thread in SDO/AIA 131~{\AA} at $\sim$16:45~UT (Figs.~\ref{fig:appear1}\&\ref{fig:appear2}), when the GOES 1--8~{\AA} flux starts to rise slowly (Fig.~\ref{fig:time}a). The brightening thread is co-spatial with a low-lying filament (Fig.~\ref{fig:appear1}d), which stays almost unperturbed during the MFR eruption (\efig~\ref{fig:filament}; \snote~1.2). The filament is aligned along the main polarity inversion line (PIL) of the active region, where a number of bald-patch (BP) candidates are identified (Fig.~\ref{fig:appear1}f; Methods). Embedded in a bundle of sheared sigmoidal loops observed in AIA 94~\AA\ (Fig.~\ref{fig:appear1}c) and almost invisible in other cool AIA channels such as 211, 193 and 171~\AA, the S-shaped thread has a mean temperature of about 7~MK with a dominant hot emission measure (EM) component at $\log T\approx$ 6.7--7.0, according to the differential emission measure (DEM) distribution sampled on the thread (labeled `S' in Fig.~\ref{fig:appear2}; Methods). The EM loci curves of six extreme-ultraviolet (EUV) channels from the thread form an envelope over the multi-peak DEM distribution (Fig.~\ref{fig:appear2}b; Methods), corroborating the DEM result. The reference DEM distribution sampled from a nearby region off the thread (labeled `R' in Fig.~\ref{fig:appear2}) shows similar cool EM components at $\log T\approx$ 5.5--6.6 but much weaker EMs at $\log T\approx$ 6.7--7.0, confirming the presence of hot plasmas in the thread. The newly appeared hot thread is later observed to continually grow into the eruptive sigmoid, hence is termed a `seed' MFR hereafter \cite{Gou2019,LiuR2020}.

\subsection*{Buildup of the MFR}

As soon as it appears, the seed MFR further extends its ``elbows'' of the inverse-S shape and grows thicker (\efig~\ref{fig:form}; \smv~3) from several Mm to $\sim\,$30~Mm within the next $\sim$35~min before the eruption, as can be seen in the stack plot made from the virtual slit S1 across the sigmoid (Fig.~\ref{fig:time}c \& \efig~\ref{fig:form}a; Methods). The growth of the seed MFR is associated with a slow rise in the GOES 1--8~\AA\ SXR flux (Fig.~\ref{fig:time}a), before the onset of intense energy release at $\sim$17:21~UT. DEM analysis evidences intense heating during the formation of the MFR (\efig~\ref{fig:dem}; \snote~1.1).

With the development of the MFR, significant pre-flare ribbon brightenings in the low atmosphere are observed by both SDO/AIA and IRIS (\efig~\ref{fig:form}; \smv~3). As the seed MFR extends to form a complete inverse-S shape, a trapezoid-shaped ribbon is spawned from the far end of the eastern ribbon (\efig~\ref{fig:form}). The spawning process is demonstrated in the stack plots of the virtual slits S2 and S3 (Fig.~\ref{fig:time}d,e \& \efig~\ref{fig:form}d; Methods). The trapezoidal ribbon shows a fast and then slow expansion, which is associated with the fast and then slow widening of the coherent sigmoid observed in AIA 131\AA\ (Fig.~\ref{fig:time}c--e). Comparing AIA 131~{\AA} and IRIS 1400~{\AA} images at about 17:12~UT (Fig.~\ref{fig:iris}a,d \& \efig~\ref{fig:form}), one can see that the trapezoidal ribbon is the footpoints of coronal loops belonging to the sigmoidal MFR. That the expansion of the trapezoidal ribbon is accompanied by the sigmoid thickening argues strongly for the development of a twisted MFR \cite{Demoulin1996fluxtube,WangWS2017}. During this phase, the instantaneous slipping motion proceeds almost perpendicularly to the overall direction of the ribbon extension, which is described as ``squirming'' motions in a previous study \cite{Dudik2016}. During the flare precursor phase, the rate of the magnetic reconnection that contributes to the MFR buildup is estimated to vary around $2\times10^{19}~\mathrm{Mx}~\mathrm{s}^{-1}$ (Fig.~\ref{fig:time}b; Methods), a few times lower than, but still on the same magnitude as, that during the MFR eruption. 

\subsection*{Ribbon evolution during eruption}

Shortly after being well developed, the MFR erupts to produce an intense X-class flare. The flare ribbon in the positive polarity region (PR) remains relatively stationary (Fig.~\ref{fig:ov}g, \efig~\ref{fig:erupt}b), probably due to its proximity to the sunspot where the strong magnetic field is relatively ``rigid''; the ribbon in the negative polarity region (NR), which is associated with the eastern foot of the MFR, evolves greatly and rapidly during the eruption (Fig.~\ref{fig:ov}g,h \& \efig~\ref{fig:erupt}; \smv~3).

From about 17:18~UT, NR's hook changes from a closed to an open morphology (Fig.~\ref{fig:iris}, \efig~\ref{fig:erupt}; \smv~3). As the ribbon moves, it sweeps the trapezoidal ribbon that encloses the eastern foot of the pre-eruption MFR. During the eruption, both ends of NR elongate rapidly southwestwards, and the whole ribbon develops into an horseshoe shape (\efig~\ref{fig:erupt}). The northern branch of NR (NR$_n$), which host negative footpoints of post-flare loops, extends rapidly along the direction of the main PIL; the motion is associated with the strong-to-weak shear transition of the post-flare arcade observed in AIA (\efig~\ref{fig:erupt}; \smv~3). Sometimes NR$_n$ itself shows two parallel branches in IRIS 1400~\AA\ (e.g., at 17:33~UT; \efig~\ref{fig:erupt}d), but they are not resolved in AIA 1600~\AA\ observations. The southern part of the horseshoe-shaped ribbon (NR$_s$) completes the hooked part of NR, which supposedly maps the footprint of the QSLs that wrap around the erupting MFR \cite{Janvier2014}.

In particular, NR$_s$ exhibits irregular, complex motions (Fig.~\ref{fig:ov}h; \smv~3). It zigzags towards southwest, in a serpentine fashion. The stack plot of the virtual slit S4 across the horseshoe-shaped ribbon (\efig~\ref{fig:erupt}d) nicely shows the zigzag motions of NR$_s$, which leave irregular trails, distinct from the rather smooth track left by NR$_n$ (Fig.~\ref{fig:time}f). As the eruption proceeds, the horseshoe shape is stretched, but its opening becomes wider, where coronal dimmings develop (Fig.~\ref{fig:iris}, \efig~\ref{fig:erupt}). During the flare decay phase, NR$_s$ fades away, and only NR$_n$ remains visible in AIA 1600~\AA\ images (Figs.~\ref{fig:ov}g\&\ref{fig:time}f). During the eruption, some bright knots are observed to distribute quasi-periodically and move along the ribbon, which may indicate that the slipping reconnection proceeds quasi-periodically \cite{LiT2015}.

The ribbon brightening in IRIS SJI 1400~\AA\ images shows three phases of temporal evolution (Fig.~\ref{fig:ribbon}), i.e., 1) the development of the trapezoidal ribbon from 16:45 to 17:18~UT, 2) the change of NR's hook from a closed to an open morphology from 17:18 to 17:26~UT, and 3) the evolution of the horseshoe-shaped NR afterwards. It is remarkable that most of the region enclosed by the trapezoidal ribbon (marked by orange dotted lines in Fig.~\ref{fig:ribbon}) that is swept during the first phase is swept again during the second phase (as shown in blue in Fig.~\ref{fig:ribbon}d), which suggests that the reconnections taking place during the second phase must have involved most of the magnetic flux in the MFR formed during the first phase. It is also remarkable that NR$_\mathrm{n}$, the ribbon segment that hosts the negative footpoints of flare loops, only appears during the third phase (Fig.~\ref{fig:ribbon}c,d). In contrast, in a different event featuring an MFR under formation during the eruption \cite{WangWS2017}, the two straight flare ribbons appear first and later extend and develop a closed hook at the far end of each ribbon, enclosing the rope's two feet.

\subsection*{Association with coronal dimming}

Enclosed by flare ribbons, two dimming regions are observed in cool AIA EUV passbands (\efig~\ref{fig:dimming}), indicative of coronal mass loss along two feet of the CME. The dimming in negative polarities as enclosed by NR is evident in AIA 335~\AA\ images, and it migrates as the MFR erupts (Fig.~\ref{fig:iris}b,c; \efig~\ref{fig:erupt}c; \smv~4). The dimming lightcurve  (Fig.~\ref{fig:time}a; Methods) shows that coronal dimming around NR starts to develop from $\sim$17:10~UT onward, when the MFR is well developed as characterized by its eastern foot being fully outlined by a trapezoidal ribbon. However, the initial dimming is not enclosed by, but located outside, the trapezoid (Fig.~\ref{fig:ribbon}e; Methods). During the eruption, in spite of strong diffraction patterns in AIA images, the dimming intensifies as shown by the 335~\AA\ lightcurve (Fig.~\ref{fig:time}a). The coronal dimming mainly occupies within the horseshoe-shaped ribbon, close to the NR$_s$ side (Fig.~\ref{fig:ribbon}f,g). As the horseshoe shape of NR is stretched, the dimming region is mainly concentrated in the opening of the horseshoe after 17:50~UT (Fig.~\ref{fig:ribbon}h), and it stays quasi-stationary thereafter during the long-duration decay phase (\smv~4). A conjugated pair of dimmings are identified on both the northern and southern sides of the post-flare arcade during the late decay phase (\efig~\ref{fig:cme}). Corresponding to the extended decreasing brightness (Fig.~\ref{fig:time}a; Methods), these dimmings continue to map the two footpoints of the CME, whose connection settles in the north-south direction (\efig~\ref{fig:cme}d), in contrast to the orientation of the pre-eruption MFR that is mainly along the east-west direction (\efig~\ref{fig:cme}a).

Combining the evolution of flare ribbon and coronal dimming (Fig.~\ref{fig:ribbon}), we conclude that the eastern foot of the MFR undergoes a drastic migration during the eruption. The centroid of magnetic fluxes through the foot is shifted by about 70$''$ during the impulsive eruption (Fig.~\ref{fig:ribbon}h; Methods). Its average migrating speed is about 26~km~s$^{-1}$, significantly faster than the typical Alfv\'en speed in the lower solar atmosphere ($\sim$10~km~s$^{-1}$) and the typical speed of photospheric flows ($\sim$1~km~s$^{-1}$). Considering the irregular and highly dynamic motions of the ribbon hook (Fig.~\ref{fig:ov}h; \smv~3), the instantaneous migrating speed would also be irregular and most likely bursty. The magnetic flux through the foot, which gives an estimation of the toroidal flux of the rope, increases from about $3.3\times10^{20}$~Mx to $7.9\times10^{20}$~Mx during the flare impulsive phase (Methods). Thus, the substantial displacement of and the associated twofold increase in magnetic flux through the foot, strongly evidence a complete restructuring of the original MFR during its development into the CME. This is radically different from the continuous deformation and drifting of flux-rope footpoints found in previous studies \cite{Aulanier&Dudik2019,Dudik2019}. 

\section*{Discussion}

To summarize, we investigated the formation and eruption of a sigmoidal MFR with coronal EUV and low-atmosphere UV observations. We observe that the coherent sigmoid builds up upon a slim S-shaped `seed' loop beneath a sheared arcade within tens of minutes during the flare precursor phase (Fig.~\ref{fig:appear1} \& \efig~\ref{fig:form}). The buildup process is featured not only by the lengthening and thickening of the S-shaped seed in the AIA 131~{\AA} passband, but also by its eastern footpoint slipping away from the core field and subsequently expanding into being enclosed by a trapezoidal ribbon in IRIS SJI~1400\AA, which argues strongly for the development of a twisted MFR. During the eruption, the trapezoidal ribbon transforms into an open hooked one. But the open hooked region has little overlapping with the original trapzoidal region (Figs.~\ref{fig:iris}\&\ref{fig:ribbon}), indicating that the eastern foot of the pre-existing MFR undergoes a drastic migration, which is associated with a twofold increase in magnetic flux through the rope's foot. These observations substantiate a complete replacement of magnetic fluxes in the original MFR by flare reconnections during the impulsive phase.

During the eruption, the ribbon hook NR$_s$, which maps the footprints of the MFR's QSL boundary, exhibits complex patterns and evolves dynamically (Fig.~\ref{fig:ov}h; \smv~3). It suggests that complicated reconnections take place at the coronal QSLs, such as between the MFR and the overlying arcade or within the MFR itself \cite{vanDriel2014,Torok2018,Aulanier&Dudik2019}. The zigzag motions of NR$_s$ and repetitive brightenings at some locations inside the hook (Fig.~\ref{fig:ribbon}; \smv~3), which are often associated with rapid changes of the hook shape (Fig.~\ref{fig:ov}h), suggest that there is an intense competition between the reconnections that turn the arcade field (anchored outside the hook) into the rope field (anchored inside the hook) and those that turn the rope field back into (post-flare) arcade field \cite{Aulanier&Dudik2019,Dudik2019}. The former reconnections build up the MFR, but the latter has the potential of eroding \cite{Aulanier&Dudik2019} or even disintegrating the original MFR. In our case, the former reconnections dominate over the latter, as indicated by the twofold increase of magnetic flux through the rope, but their competition drives the migration of the pre-eruption MFR's eastern foot to a completely new place during the eruption (Fig.~\ref{fig:ribbon}). %A further migration is manifested in the shifting of the conjugated pair of coronal dimmings (\efig~\ref{fig:cme}). 

The peak reconnection rate during the eruption is as high as $\sim1.0\times10^{20}$~Mx~s$^{-1}$ (Fig.~\ref{fig:time}b; Methods). In comparison with the toroidal flux through the flux rope, which is in the order of $10^{20}$~Mx, the total reconnected flux is about $9.5\times10^{21}$~Mx (Methods; until 19:00~UT during the flare decay phase), which falls into top $\sim$5\% of poloidal fluxes of MCs in statistical studies \cite{QiuJ2007,WangYM2015}. In addition, the total reconnected flux during the flare is as high as $\sim$66\% of the unsigned magnetic fluxes in the active region, in the context of statistical results showing that no more than 50\% of the total active-region flux is involved in the largest flares \cite{Kazachenko2017,Tschernitz2018}.

This on-disk sigmoid event provides a top view of a sigmoid growing continually from a hot inverse-S shaped thread (\efig~\ref{fig:form}), which complements the side view of a limb eruption in a previous study \cite{Gou2019} to substantiate the scenario that a large-scale MFR may build up from a small-scale seed rope, with the seed formation marking the sudden commencement of the transition from a sheared arcade to a CME MFR \cite{LiuR2020,Gou2019,Jiang2021} (\snote~1.1). Since the footpoints of the pre-eruption MFR are distinct from those of the CME (Fig.~\ref{fig:ribbon}, \efig~\ref{fig:cme}), the eruptive scenario is different from the standard picture, in which flare reconnections build up a pre-existing MFR by turning untwisted arcade field lines into twisted field lines wrapping around the rope \cite{Lin&Forbes2000,Aulanier2012,Janvier2013}. In contrast, flare reconnections in the current observation that completely replace the magnetic flux of the pre-existing MFR are manifested as the highly dynamic motions of the hooked ribbon and the long-distance footpoint migration, which are also different from the continuous deformation and gradual drifting of the MFR footpoints as predicted by idealized 3D MHD simulations \cite{Janvier2013,Aulanier&Dudik2019}.

Although these observations were neither anticipated by the standard picture nor reproduced in numerical simulations, they must be a consequence of 3D magnetic reconnection that is inherently built into the 3D extension of the standard model\cite{Aulanier2012,Aulanier2013,Janvier2013,Janvier2014}. As an important topological component in the 3D model, the hooked ribbon demonstrates complex changes and motions, which reveals new details about 3D magnetic reconnections that reshape the pre-existing MFR into the CME. Such a significant restructuring of pre-existent MFRs during eruptions would add another layer of difficulty to forecasting space weather, but may shed light on complex ejecta in interplanetary space (\snote~1.3).

\section*{Methods}

\subsection*{Instruments and Data}

The Atmospheric Imaging Assembly \cite{Lemen2012} (AIA) onboard the Solar Dynamics Observatory \cite{Pesnell2012} (SDO) takes full-disk images of the Sun around the clock. The seven EUV channels, i.e., 131~\AA\ (primarily from \ion{Fe}{21} emission line, $\log T=7.05$), 94~\AA\ (\ion{Fe}{18}, $\log T=6.85$), 335~\AA\ (\ion{Fe}{16}, $\log T=6.45$), 193~\AA\ (primarily \ion{Fe}{12}, $\log T=6.2$), 211~\AA\ (\ion{Fe}{14}, $\log T=6.3$), 171~\AA\ (\ion{Fe}{9}, $\log T=5.85$), and 304~\AA\ (\ion{He}{2}, $\log T=4.7$), obtain images with a spatial resolution of $0.''6$ and a temporal cadence of 12~s. Two UV channels, 1600~\AA\ (\ion{C}{4} line and continuum emission, $\log T=5.0$) and 1700~\AA\ (continuum), obtain images with a temporal cadence of 24~s. AIA data are processed to level 1.5, and the heliographic maps are all rotated to 16:00~UT to correct the solar rotation.

The Helioseismic and Magnetic Imager \cite{Schou2012} (HMI) onboard SDO measures the magnetic fields on the solar photosphere. We use HMI line of sight magnetograms to compare with heliographic observations by AIA and IRIS. To investigate the detailed magnetic configuration of the active region, we use the HMI active region patches (SHARP) vector magnetograms, which are taken every 12 minutes and remapped with the cylindrical equal area (CEA) projection, with a pixel scale of about 0.36~Mm. At the PIL of the active region which is identified by the contour of $B_z =0$, we search for the existence of bald patches (BPs; Fig.~\ref{fig:appear1}f) by employing the criterion $(B_\bot \cdot \nabla_\bot B_z)|_{\rm{PIL}} > 0$ \cite{Titov1993,Titov&Demoulin1999}, where magnetic field vectors are directed from negative to positive polarities and the field line is tangential to the photosphere and concave upward. BPs are topological structures known to be favorable for the formation of current sheets and therefore for the occurrence of magnetic reconnection (see also \snote~1.2). 

The Interface Region Imaging Spectrograph (IRIS)\cite{DePontieu2014} captures the formation and eruption phase of the event before 17:58~UT. The Slit Jaw Imager (SJI) obtains images with a 19-s cadence and a spatial resolution as high as $\tfrac16''$ over a field of view (FOV) of $119'' \times 119''$, which fortuitously covers the eastern foot of the MFR (\smv~3). IRIS level-2 data of SJI Si IV 1400~\AA\ passband ($\log T\approx$ 4.9, forming in the transition region) are used in the study. The superior spatiotemporal resolution of IRIS is key to this study; e.g., it registers brightening in 1400~{\AA} around the eastern footpoint of the seed MFR as soon as it appears in EUV at 16:45~UT (Fig.~\ref{fig:appear1}e), suggestive of simultaneous heating in the low atmosphere. However, the brightening is too thin to be clearly resolved by AIA ultraviolet (UV) channels during this early phase. The subsequent footpoint evolution of the seed MFR is resolved by both IRIS and SDO/AIA, but the morphology and evolution of flare ribbons are better resolved by IRIS both temporally and spatially. IRIS observations hence serve as a reference and guide for our identification of flare ribbons observed by AIA. Together, IRIS and AIA observations make it possible to estimate the pre-flare reconnection rate (see below), which is rarely given in the literature.

To characterize the temporal evolution, we place four virtual slits (S1--S4; \efigs~\ref{fig:form}\&\ref{fig:erupt}) in SDO/AIA and IRIS images and generate the time--distance stack plots (Fig.~\ref{fig:time}c--f).  S1 is put across the sigmoid in SDO/AIA 131~\AA\ images (\efig~\ref{fig:form}a), with a length of 76~Mm, averaged over 6 AIA pixels. S2 and S3 are across the trapezoid-shaped ribbon in IRIS 1400~\AA\ (\efig~\ref{fig:form}d), with a length of 47 and 59~Mm, respectively, averaged over 10 IRIS pixels. S4 is across the horseshoe-shaped ribbon in IRIS 1400~\AA\ (\efig~\ref{fig:erupt}d), with a length of 45~Mm, averaged over 10 pixels. Stack plots of S2--S4 after 17:58~UT are generated from SDO/AIA 1600~\AA\ images at the same locations, when there are no IRIS observations.

The Geostationary Operational Environmental Satellite (GOES) spacecraft operates in geosynchronous orbit and measures the solar soft X-ray flux in units of $\text{W}\,\text{m}^{-2}$. The classification of solar flares uses the letters A, B, C, M, or X according to the peak flux in 1--8~{\AA} as measured by GOES, with each letter representing a 10-fold increase in flux.

\subsection*{DEM analysis}

SDO/AIA's six optically thin EUV passbands (except for 304~\AA) are used for differential emission measure (DEM) analysis to study the temperature characteristics. The AIA data are further processed to level 1.6 before being fed into the DEM code. Here we used the sparse inversion code \cite{Cheung2015}, which has been further modified to optimize solutions for hot plasmas above a few MK\cite{SuY2018}. DEMs are calculated in the temperature range of $\log T$ = 5.5--7.5 with an interval of $\Delta\log T$ = 0.05, and a DEM-weighted mean temperature, $\langle T\rangle=(\sum DEM(T) \times T\Delta T)/(\sum DEM(T) \Delta T)$, is also derived \cite{Gou2015}. DEM uncertainties are obtained by 250 times of Monte Carlo simulations. We also computed the EM loci curves ($\mathrm{EM}_\lambda(T)=I_\text{obs}/\epsilon_\lambda(T)$) of a sampled region using the observed intensities $I_\text{obs}$ and temperature response functions $\epsilon_\lambda(T)$ from the six AIA channels.

\subsection*{Identification of ribbon and coronal dimming}

To characterize the ribbon evolution, we identify the low-atmosphere ribbon by counting brightened pixels in SDO/AIA UV channels (Fig.~\ref{fig:ov}g) and IRIS SJI 1400~\AA\ passbands (Fig.~\ref{fig:ribbon}), respectively. During the precursor phase, flare brightening in AIA 1600~\AA\ is too weak to be clearly distinguished from chromospheric plages, thus we use the ratio of AIA 1600~\AA\ and 1700~\AA\ passbands to enhance the ribbon brightening \cite{Dudik2016} (\efig~\ref{fig:form}; \smv~3). After the flare onset at 17:21~UT, only AIA 1600~\AA\ passband is used. To best identify the flare ribbons, different threshold values are set during different evolutionary phases.

The identified ribbon pixels in SDO/AIA are used to calculate the reconnection flux (Fig.~\ref{fig:time}b), after being projected onto a pre-flare SDO/HMI SHARP $B_z$ map with the CEA projection. We measured the instantaneous reconnection flux by summing up magnetic fluxes swept by newly brightened ribbon pixels, and estimated the uncertainties by the difference of the measured magnetic flux in positive and negative polarities. Integrating the instantaneous reconnection flux with respect to time, we obtained the accumulative reconnection flux.

Compared to AIA, IRIS is superior in spatial resolution and dynamic range, though it has a limited field of view mainly covering the ribbon of negative polarity  (\efigs~\ref{fig:form} \& \ref{fig:erupt}, \smv~3). We manually traced the bright ribbon fronts observed in IRIS 1400~\AA\ at selected time instants to highlight the complex patterns and irregular motions (Fig.~\ref{fig:ov}h).

Coronal dimming around the ribbon NR are evident and migrate in SDO/AIA 335~\AA\ passband during the MFR eruption (\efig~\ref{fig:erupt}; \smv~4). To study the dimming evolution, we identified dimmings as the pixels whose brightness decreases compared to that before the eruption (we refer to 16:20~UT for the event; Fig.~\ref{fig:ribbon}e--h), and took the average brightness to obtain the 335~\AA\ dimming lightcurve (Fig.~\ref{fig:time}a). For the conjugated pair of post-eruption dimmings detected at the footpoints of large-scale overlying loops (\efig~\ref{fig:cme}d), we sampled intensities at two representative locations in SDO/AIA 171~\AA\ (`DS' and `DN' in \efig~\ref{fig:cme}d, $10''\times10''$) to show their temporal evolution (Fig.~\ref{fig:time}a).

To quantitatively study the footpoint migration, we compare two time instants before and after the eruption as an illustration, when the eastern foot of the MFR is well defined; i.e., at about 17:18~UT when it is enclosed by a trapezoidal ribbon (Fig.~\ref{fig:ribbon}a), and after 17:50~UT when it is marked by the coronal dimming that is concentrated in the opening of the ribbon hook (e.g., Fig.~\ref{fig:ribbon}h). One can see that the centroid of magnetic fluxes within the foot migrates from about [-167$''$, 111$''$] to [-108$''$, 76$''$] (Fig.~\ref{fig:ribbon}h), which is shifted by about 70$''$ with an average migrating speed of about 26~km~s$^{-1}$. The magnetic flux through the foot changes from about $3.3\times10^{20}$~Mx to $7.9\times10^{20}$~Mx accordingly, which increases nearly twofold. While the foot boundary is difficult to determine at other times during the eruption, the substantial displacement of the foot and the increase of magnetic flux through it strongly evidence a complete restructuring of the original MFR during its development into the CME.

\subsection*{Data availability}
The data used in the study are publicly available for download from the corresponding mission archives. SDO data are available at \url{http://jsoc.stanford.edu/ajax/lookdata.html}; IRIS data are available at \url{https://iris.lmsal.com/data.html}.

\subsection*{Acknowledgments} 
We acknowledge SDO and IRIS teams for the science data. IRIS is a NASA small explorer mission developed and operated by LMSAL with mission operations executed at NASA Ames Research Center and major contributions to downlink communications funded by ESA and the Norwegian Space Centre. T.G. thanks the BBSO staff for providing H$\alpha$ data. T.G., R.L., W.W., M.X., and Y.M. acknowledge the support from National Natural Science Foundation of China (NSFC 42188101, 42274204, 11903032, 11925302, 12003032) and the Strategic Priority Program of the Chinese Academy of Sciences (XDB41030100). A.M.V. acknowledges the Austria Science Fund (FWF): project I-4555N. T.L. acknowledges the support from NSFC 12222306. T.G. also acknowledges the support from CAS Key Laboratory of Solar Activity. 

\subsection*{Author contributions} 
T.G. and R.L. led the study and analysis, interpreted the data, and wrote the manuscript. A.M.V. discussed the analysis and contributed to the interpretation, conclusion and writing of the manuscript. B.Z. led the in-situ data analysis and discussed the interpretation. T.L. and Y.W. discussed the results and contributed to the interpretation. W.W. and M.X. contributed to the SDO and in-situ data analyses. All authors participated in the discussion and contributed to finalizing the manuscript.

\subsection*{Competing interests} The authors declare no competing interests.

%\section*{Figures}

\clearpage
\renewcommand{\figurename}{\textbf{Fig.}}
\setcounter{figure}{0}
\spacing{1}

\begin{figure}[htbp]
	\centering
	\includegraphics[width=\textwidth]{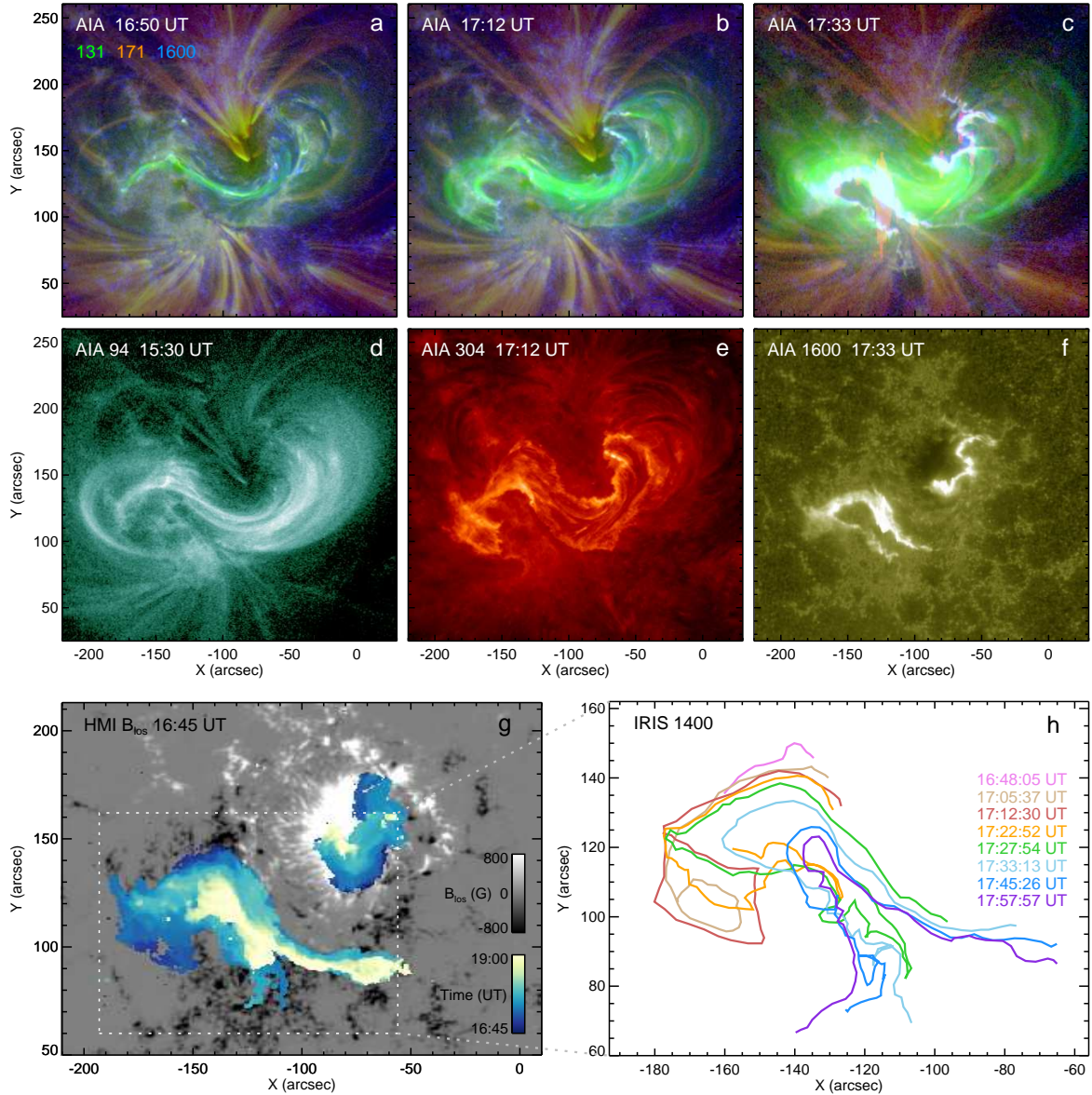}
	\caption{\small \textbf{Overview of 2014 September 10 sigmoid event.} \textbf{(a--c)} SDO/AIA 131~\AA\ (green), 171~\AA\ (orange) and 1600~\AA\ (blue) composite images during different evolutionary phases, showing the seed MFR beneath large-scale active-region loops, the coherent sigmoidal MFR before eruption, post-flare loops and flare ribbons after the eruption, respectively. \textbf{(d)} SDO/AIA 94~\AA\ image about two hours before the eruption, showing sheared S-shaped loops. \textbf{(e)} SDO/AIA 304~\AA\ image observed at the same time with panel (b), showing the trapezoid-shaped ribbon co-spatial with the eastern foot of the sigmoid. \textbf{(f)} SDO/AIA 1600~\AA\ image observed at the same time with panel (c), featuring the flare ribbon during the eruption. \textbf{(g)} SDO/HMI line of sight magnetogram before the flare, superposed by color-coded temporal evolution of ribbon brightenings identified in AIA UV passbands (Methods). \textbf{(h)} A schematic zooming in on representative ribbon fronts in negative polarities, which are manually traced from IRIS SJI 1400~\AA\ images at selected time instants (Methods).}
	\label{fig:ov}
\end{figure}

\begin{figure}[htbp]
	\centering
	\includegraphics[width=\textwidth]{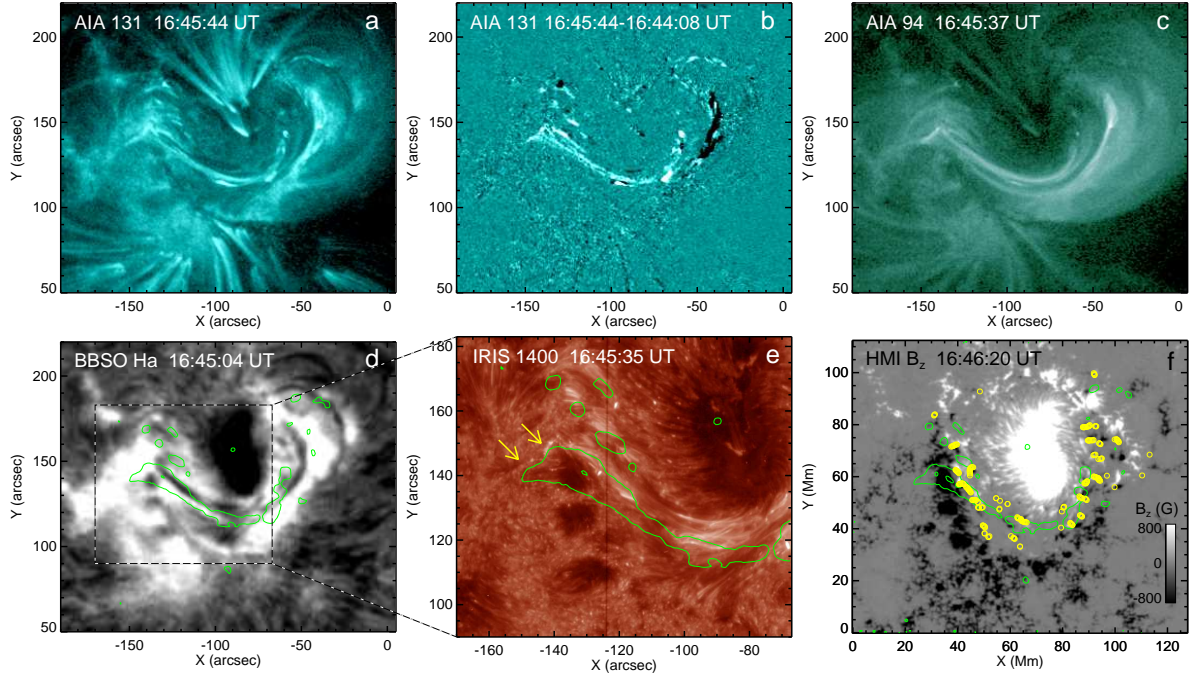}
	\caption{\small \textbf{Appearance of the seed MFR at 16:45~UT.} \textbf{(a--c)} SDO/AIA 131~\AA, its difference, and 94~\AA\ images, featuring the S-shaped thread beneath the sheared arcade. \textbf{(d--e)} Big Bear Solar Observatory (BBSO) H$\alpha$ line center and IRIS SJI 1400~\AA\ images. The dotted rectangle in panel (d) indicates the FOV of panel (e).  Green contours superimposed in panels (d \& e) are taken from the S-shaped thread observed in AIA 131~\AA . Yellow arrows in panel (e) mark the IRIS 1400~\AA\ brightenings associated with the appearance of the S-shaped thread in AIA 131~\AA. \textbf{(f)} SDO/HMI SHARP B$_z$ map with CEA projection. Yellow symbols indicate BP candidates at the PIL of the active region (Methods). Superimposed green contours are the same as in panels (d \& e) except being remapped with CEA projection.}
	\label{fig:appear1}
\end{figure}

\begin{figure}[htbp]
	\centering
	\includegraphics[width=\textwidth]{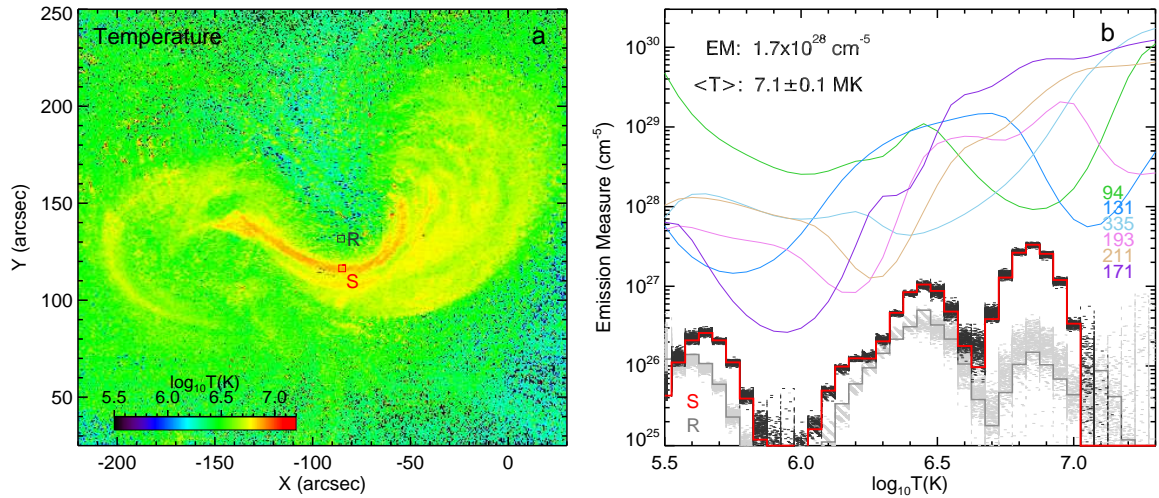}
	\caption{\small \textbf{Temperature structure of the seed MFR.} \textbf{(a)} Map of DEM-weighted mean temperature as scaled by the color bar (Methods). \textbf{(b)} EM loci curves of six AIA EUV channels (colored lines; Methods) and DEM distribution (the red thick curve) sampled from a region (4$\times$4~pixel$^2$) on the S-shaped thread (indicated by the red box `S' in panel a). The reference DEM distribution from a nearby region (`R' in panel a) is shown by the gray thick curve. Black and light gray curves give their Monte Carlo simulations (Methods). Annotated are the total EM and mean temperature of the thread sample `S'.}
	\label{fig:appear2}
\end{figure}

\begin{figure}[htbp]
	\centering
	\includegraphics[width=\textwidth]{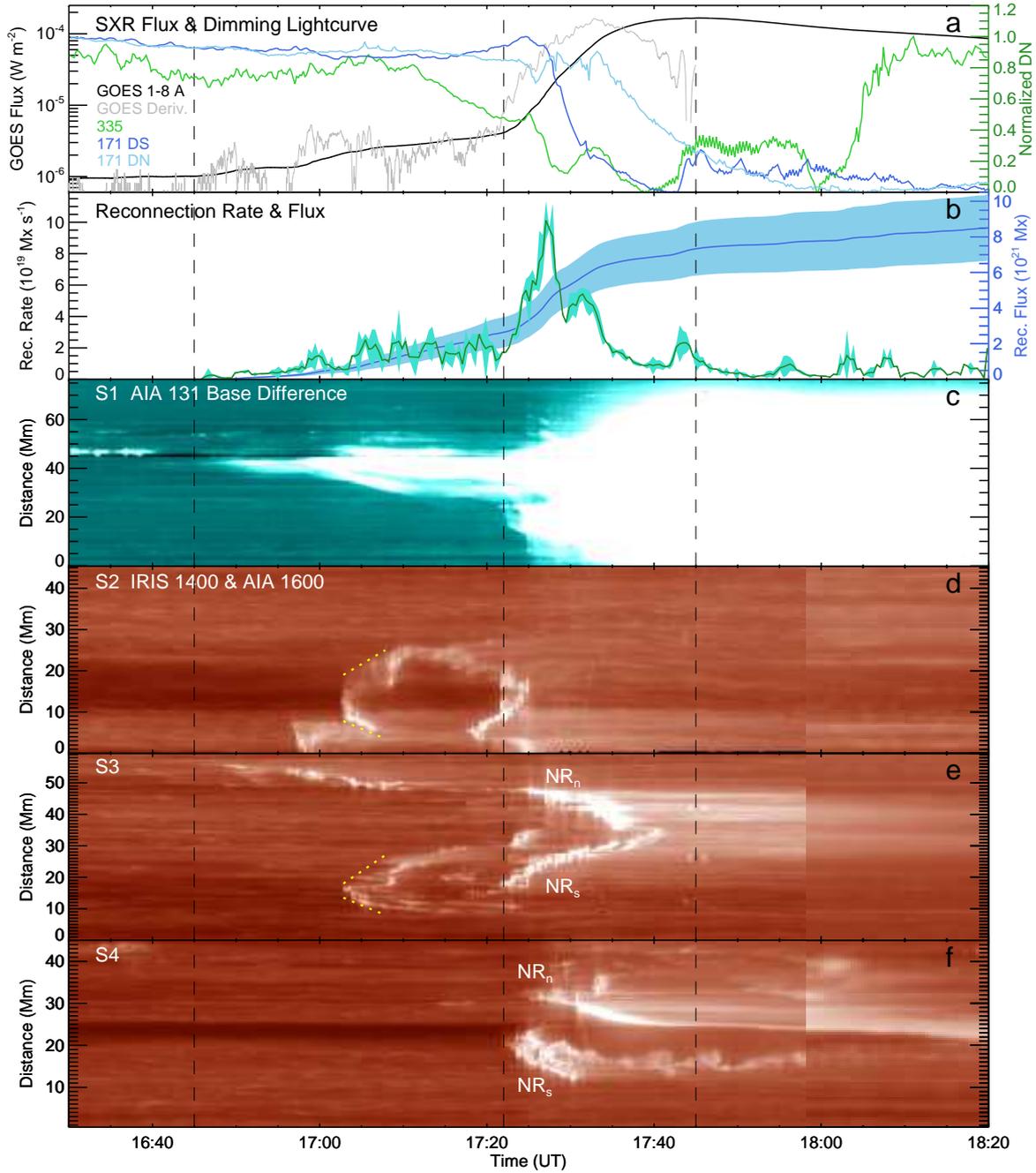}
	\caption{\small \textbf{Temporal evolution of the event.} \textbf{(a)} GOES 1--8~\AA\ SXR flux (black solid line; scaled by the left y-axis), its time derivative (gray line; scaled by an arbitrary y-axis), and normalized dimming lightcurves obtained from AIA 335~\AA\ and 171~\AA\ (colored lines; scaled by the right y-axis; see Methods and \efig~\ref{fig:cme}d). \textbf{(b)} Magnetic reconnection rate (green; scaled by the left y-axis) and the accumulative reconnection flux (blue; scaled by the right y-axis; Methods). Light green and blue shades are corresponding uncertainties. \textbf{(c--f)} Stack plots made from four virtual slits, S1--S4, as denoted in \efigs~\ref{fig:form} \& \ref{fig:erupt} (Methods). Yellow dotted lines in panels (d,e) mark the tracks left by the fast expansion of the trapezoidal ribbon. NR$_n$ \& NR$_s$ in panels (e,f) mark the tracks left by the two branches of the negative ribbon (see \efig~\ref{fig:erupt}d). The three vertical dashed lines mark successively when the seed MFR appears (16:45 UT) and when the GOES X1.6-class flare starts (17:21 UT) and peaks (17:45 UT).}
	\label{fig:time}
\end{figure}

\begin{figure}[htbp]
	\centering
	\includegraphics[width=\textwidth]{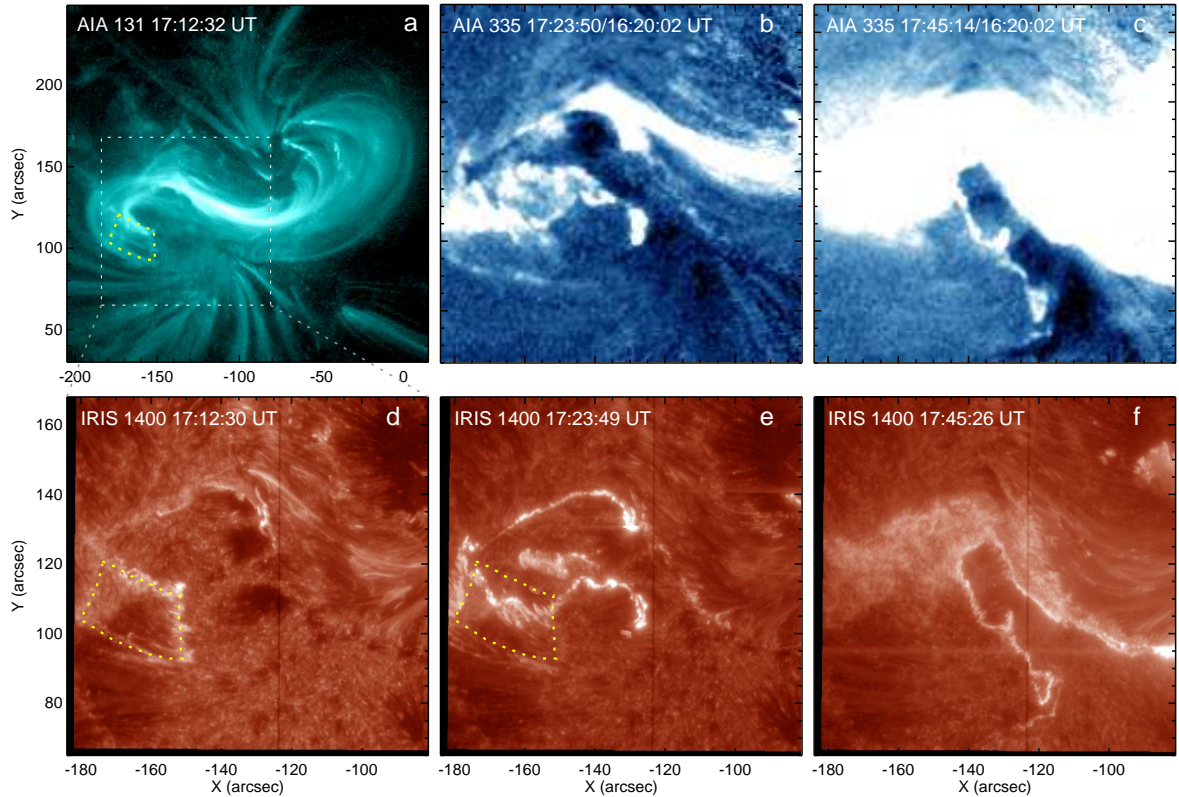}
	\caption{\small \textbf{Footpoint migration during the MFR eruption.} \textbf{(a)} SDO/AIA 131~\AA\ observation showing the pre-eruption sigmoid. The white dotted rectangle indicates the zoom-in FOV of panels (b--f). \textbf{(b,c)} SDO/AIA 335~\AA\ difference images featuring the coronal dimming during the flux rope eruption. \textbf{(d--f)} IRIS SJI 1400~\AA\ images observed at the same time with the AIA images in panels (a--c), respectively, showing the dynamic evolution of the flare ribbon. The yellow dotted lines in panels (a,d,e) denote the location of the trapezoidal ribbon in panel (d) (see also \efigs~\ref{fig:form}\&\ref{fig:erupt}), which is co-spatial with the eastern foot of the sigmoid in panel (a).}
	\label{fig:iris}
\end{figure}

\begin{figure}[htbp]
	\centering
	\includegraphics[width=\textwidth]{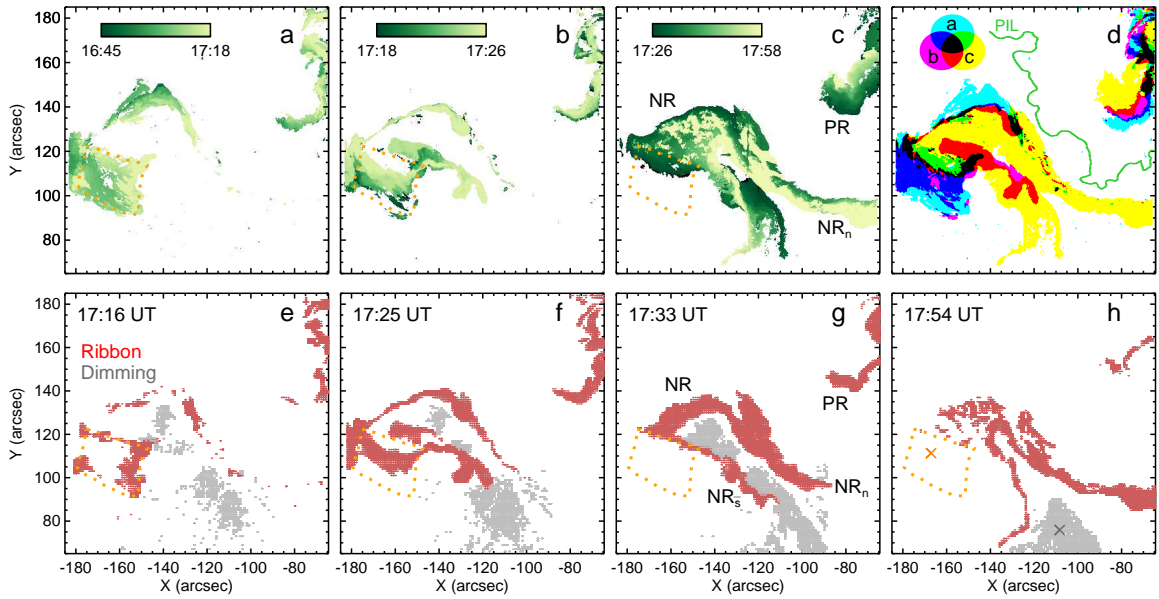}
	\caption{\small \textbf{Evolution of flare ribbons in relation to coronal dimmings.} \textbf{(a--c)} Color-coded temporal evolution of the ribbon brightening as identified in IRIS SJI 1400~\AA\ during three successive time intervals (Methods). \textbf{(d)} Synthesized ribbons from panels (a--c). The ribbon-swept region in each individual time interval is marked by cyan, magenta, and yellow colors, respectively. Regions that are swept over more than once, i.e., by ribbons developed in different time intervals, are hence shown in blended colors. The green curve shows the main PIL of the active region. \textbf{(e--h)} Snapshots of IRIS SJI 1400~\AA\ ribbon and SDO/AIA 335~\AA\ dimming, which are obtained by summing up brightened and dimmed pixels over a $\sim$1-min interval around the specified time instants, respectively. Dotted orange lines denote the trapezoidal ribbon in panel (a). The orange and dark cross symbols in panel (h) indicate the centroids of magnetic fluxes in the trapezoidal ribbon and in the dimming region, respectively (Methods).}
	\label{fig:ribbon}
\end{figure}

\clearpage
\renewcommand{\figurename}{\textbf{Extended Data Fig.}}
\setcounter{figure}{0}
\spacing{1}

\begin{figure}[htbp]
	\centering
	\includegraphics[width=\textwidth]{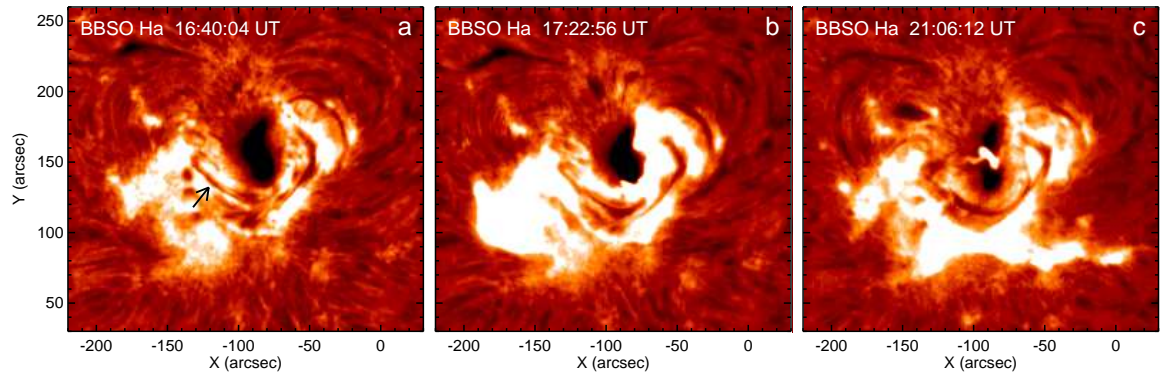}
	\caption{\small \textbf{H$\alpha$ observation of the filament from the Big Bear Solar Observatory (BBSO).} Panels (a--c) show the H$\alpha$ line center images before, during, and after the sigmoid eruption, respectively. The arrow in panel (a) denotes the filament aligned with the main PIL of the active region.}
	\label{fig:filament}
\end{figure}

\begin{figure}[htbp]
	\centering
	\includegraphics[width=\textwidth]{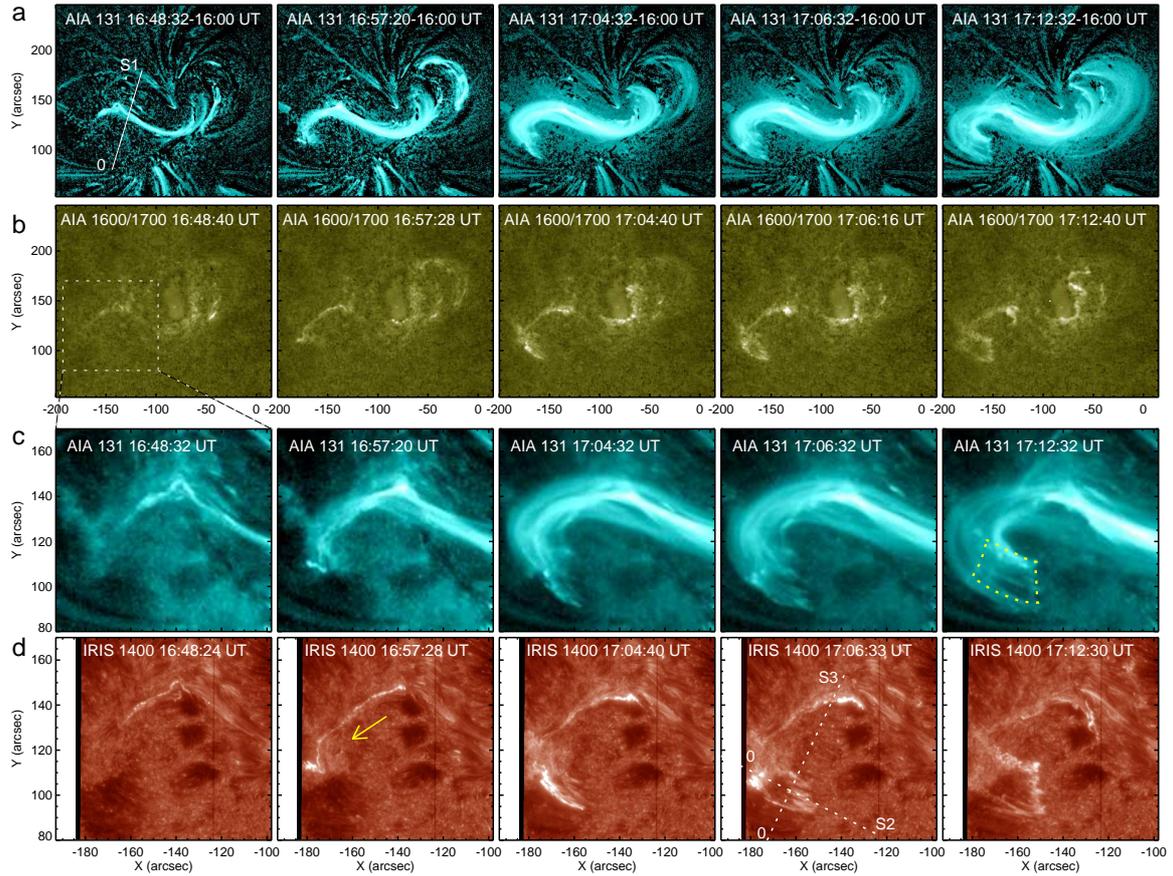}
	\caption{\small \textbf{MFR buildup during the flare precursor phase.} From the top to bottom are SDO/AIA 131~\AA\ base-difference images, ratio of SDO/AIA 1600~\AA\ and 1700~\AA\ images, SDO/AIA 131~\AA\ images zooming in on the eastern foot of the sigmoid, and simultaneous IRIS SJI 1400~\AA\ images (Methods). The white dotted rectangle in panel (b) indicates the FOV of panels (c \& d). Three white lines in panels (a \& d) indicate the virtual slits S1--S3 for the stack plots in Fig.~\ref{fig:time}(c--e) (Methods), whose starting points are labeled as `0'. The yellow arrow in the IRIS image indicates the extending direction of the ribbon. Yellow dotted lines in the AIA 131~\AA\ image at 17:12~UT denote the location of the trapezoidal ribbon observed in IRIS 1400~\AA\ at the same time (see also Fig.~\ref{fig:iris}a\&d).}
	\label{fig:form}
\end{figure}

\begin{figure}[htbp]
	\centering
	\includegraphics[width=\textwidth]{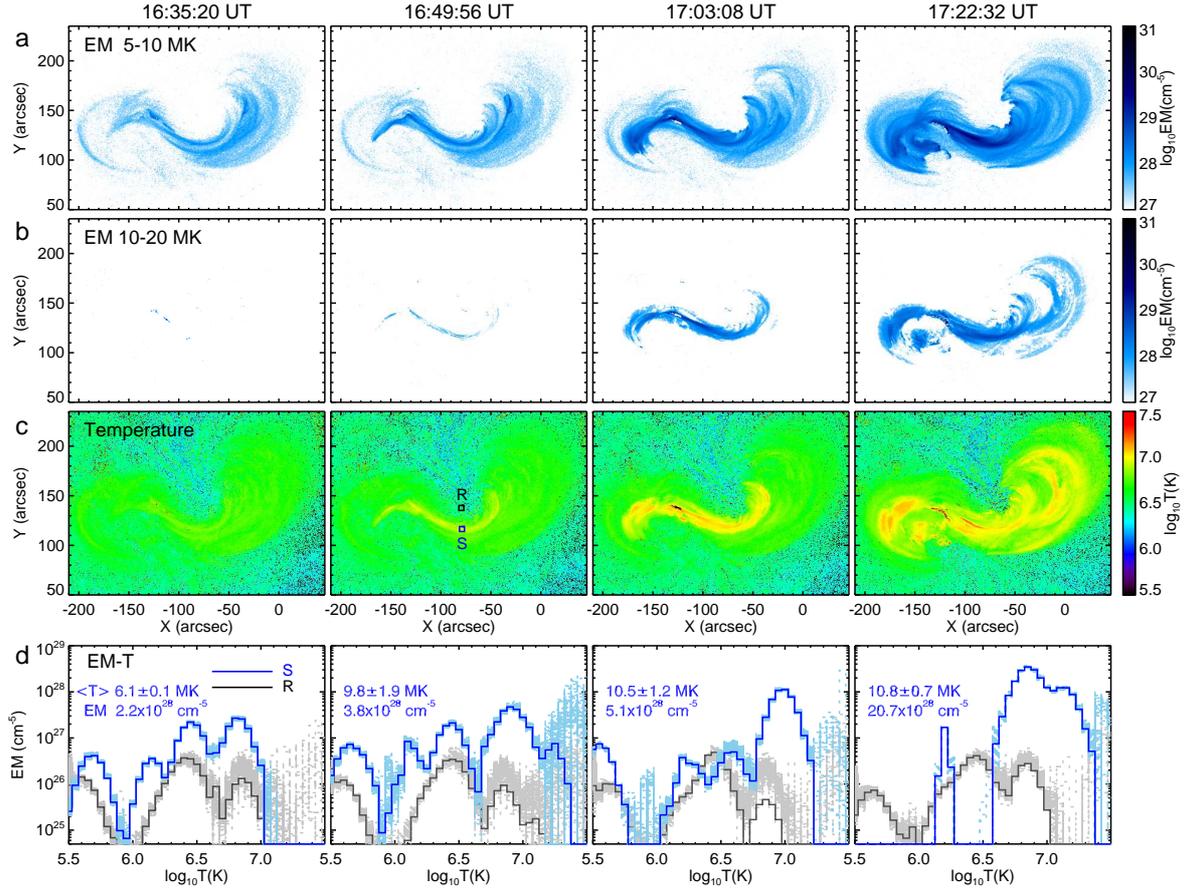}
	\caption{\small \textbf{DEM results in different evolutionary phases.} \textbf{(a,b)} EM maps in temperature ranges of 5--10 and 10--20~MK, respectively. \textbf{(c)} Maps of the DEM-weighted mean temperature (Methods). \textbf{(d)} DEM distributions sampled from two locations (4$\times$4~pixels$^2$), one on the sigmoid (`S') and the other from nearby as a reference (`R'), in blue and black, respectively. Light blue and gray curves give 250 times of Monte Carlo simulations as an estimation of the DEM uncertainty (Methods). The mean temperature and total EM of the sigmoid sample `S' are annotated.}
	\label{fig:dem}
\end{figure}

\begin{figure}[htbp]
	\centering
	\includegraphics[width=\textwidth]{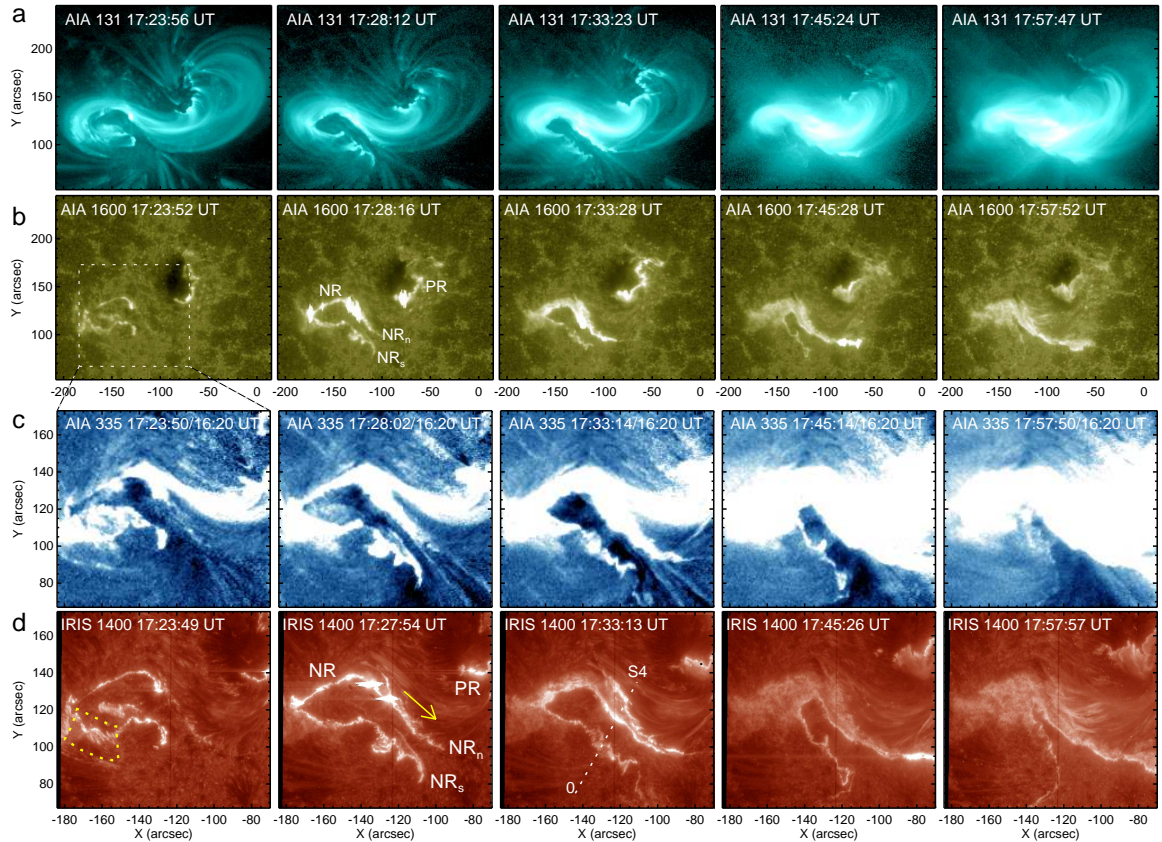}
	\caption{\small \textbf{SDO/AIA and IRIS observations during the MFR eruption phase.} From the top to bottom are SDO/AIA 131~\AA, 1600~\AA, 335~\AA\ base ratio, and IRIS SJI 1400~\AA\ images. The white dotted rectangle in panel (b) indicates the FOV of panels (c \& d). The yellow dotted lines in IRIS 1400~\AA\ are the same as those in Fig.~\ref{fig:iris} \& \efig~\ref{fig:form}. The yellow arrow indicates the extending direction of the negative ribbon NR$_n$. The white dotted line indicates the virtual slit S4 for the stack plot in Fig.~\ref{fig:time}f (Methods), with its starting point labeled by `0'.}
	\label{fig:erupt}
\end{figure}

\begin{figure}[htbp]
	\centering
	\includegraphics[width=\textwidth]{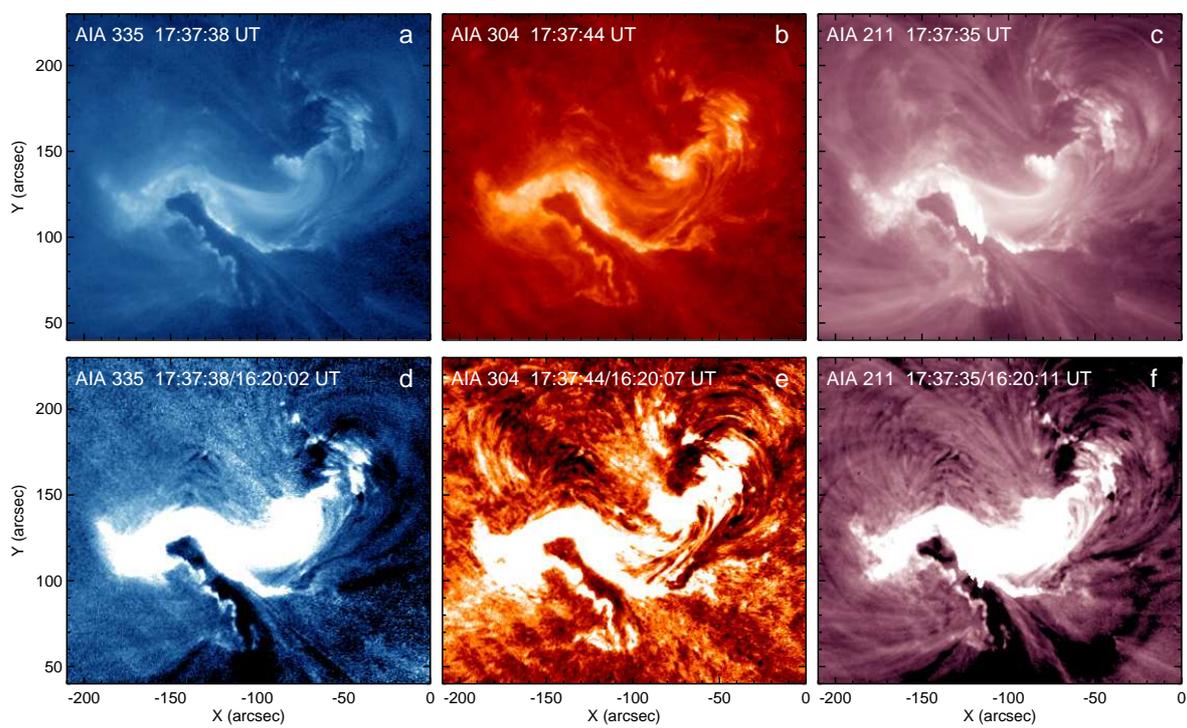}
	\caption{\small \textbf{Coronal dimmings during the MFR eruption.} Panels (a--c) show the SDO/AIA 335~\AA, 304~\AA, and 211~\AA\ images, and panels (d--f) show corresponding base-ratio images.}
	\label{fig:dimming}
\end{figure}

\clearpage
\begin{figure}[htbp]
	\centering
	\includegraphics[width=\textwidth]{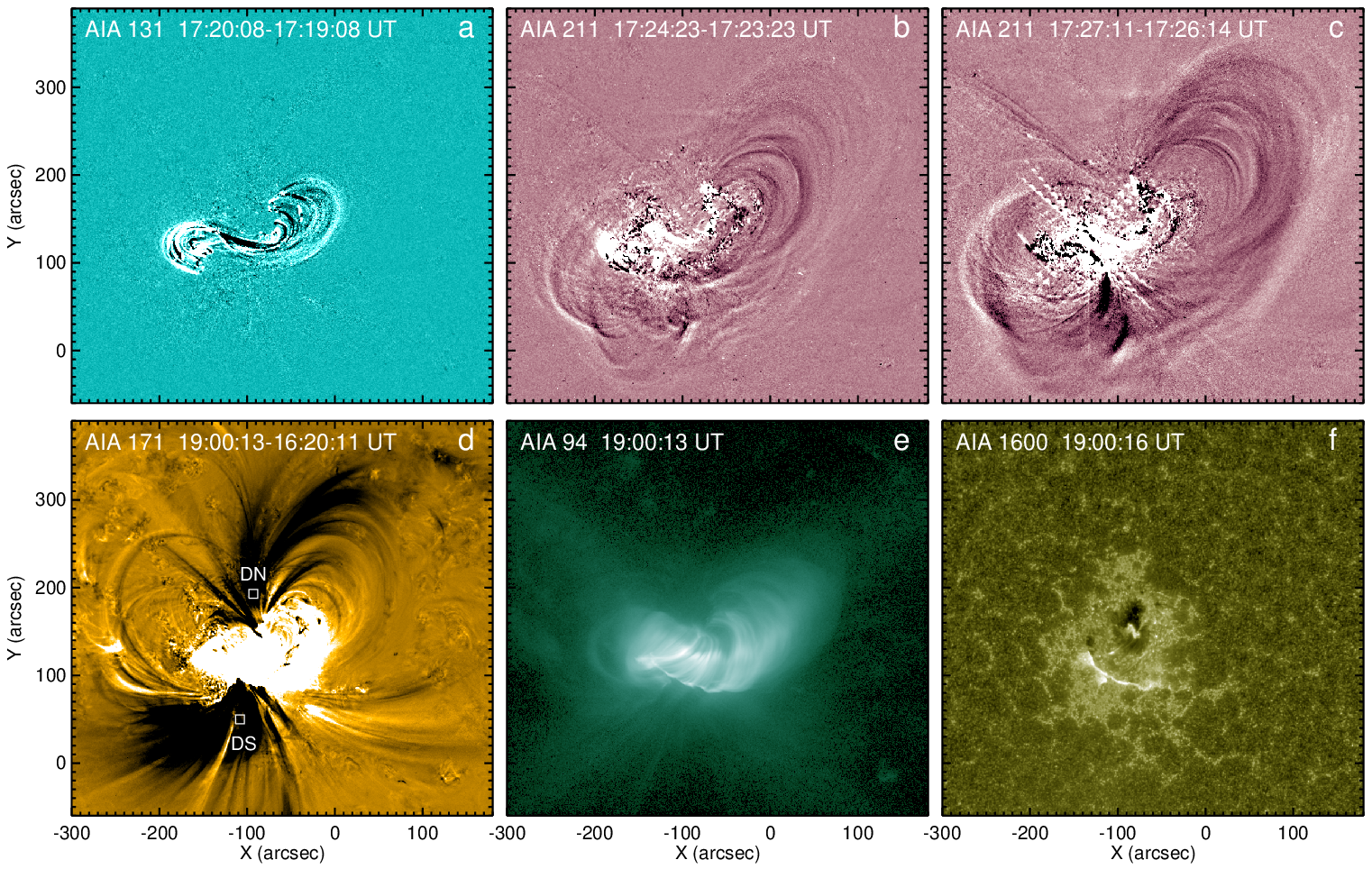}
	\caption{\small \textbf{SDO/AIA observations of the sigmoid eruption.} \textbf{(a--c)} AIA 131~\AA\ and 211~\AA\ running-difference images showing the pre-eruption MFR and the erupting CME. \textbf{(d--f)} AIA observations of coronal dimmings in 171~\AA\ base-difference image, post-flare loops in 94~\AA, and flare ribbons in 1600~\AA, during the flare decay phase. Two boxes in panel (d) indicate the regions to obtain 171~\AA\ dimming lightcurves in Fig.~\ref{fig:time}a (Methods).}
	\label{fig:cme}
\end{figure}

\clearpage

\newpage
\setcounter{page}{1}

\spacing{1.25}

\begin{center}
    
\LARGE 

\textbf{Supplementary Information} \\
	    \hspace*{\fill} \\ \hspace*{\fill} \\
        Complete replacement of magnetic flux in a flux rope \\during a coronal mass ejection\\

\hspace*{\fill}

\Large
Tingyu Gou$^{1*}$, Rui Liu$^{1,2*}$, Astrid M. Veronig$^{3}$, Bin Zhuang$^{4}$,\\
Ting Li$^{5,6}$, Wensi Wang$^{1,7}$, Mengjiao Xu$^{1,8}$, Yuming Wang$^{1,2,8}$ 

\end{center}

\hspace*{\fill}

\begin{affiliations}
	%\small
	\item School of Earth and Space Sciences/CAS Center for Excellence in Comparative Planetology/CAS Key Laboratory of Geospace Environment, University of Science and Technology of China, Hefei 230026, China
	\item Collaborative Innovation Center of Astronautical Science and Technology, Harbin, China
	\item Institute of Physics \& Kanzelh\"{o}he Observatory for Solar and Environmental Research, University of Graz, 8010 Graz, Austria
	\item Institute for the Study of Earth, Oceans, and Space, University of New Hampshire, Durham, NH 03824, USA
	\item CAS Key Laboratory of Solar Activity, National Astronomical Observatories, Beijing 100101, China 
	\item School of Astronomy and Space Science, University of Chinese Academy of Sciences, Beijing 100049, China
	\item Mengcheng National Geophysical Observatory, University of Science and Technology of China, Mengcheng 233500, China
	\item Deep Space Exploration Laboratory, Hefei 230026, China \\
	*E-mail: tygou@ustc.edu.cn; rliu@ustc.edu.cn
\end{affiliations}

\hspace{2pt}

\clearpage

%\newpage
%\setcounter{page}{1}

\tableofcontents

\clearpage

\section{Supplementary Notes}

\subsection{Formation of the Sigmoidal Flux Rope}

We observe in this event that a coherent sigmoidal flux rope builds up upon a slim S-shaped seed, which appears beneath a sheared arcade in the active region, accompanied by simultaneous ribbon-like brightenings in the low solar atmosphere (Fig.~\ref{fig:appear1}). The flux rope grows continually from the hot S-shaped seed within 30--40 minutes during the flare precursor phase (\efig~\ref{fig:form}). The buildup process manifests itself in the continual lengthening and thickening of the S-shaped loop in the AIA 131\AA\ passband, and in its eastern footpoint slipping away from the core field and subsequently expanding into a trapezoidal ribbon in IRIS SJI 1400\AA\ (\efig~\ref{fig:form}; \smv~3), which argues strongly for the development of a twisted magnetic flux rope \cite{Demoulin1996fluxtube,WangWS2017}. While tether-cutting reconnections \cite{vanBallegooijen1989,Moore2001} may help produce the S-shaped thread, as evidenced by the persistent flux cancellation \cite{Green2011} in this decayed active region \cite{ChengX2015}, the subsequent growth of the sigmoid is mainly featured by the footpoint slippage (\efig~\ref{fig:form}; \smv~3), which suggests that 3D slipping reconnections \cite{Aulanier2007} play an important role.

We studied the temperature characteristics of the sigmoid and its evolution via DEM analysis. The results (\efig~\ref{fig:dem}; Methods) show that the flux rope structure has a mean temperature as high as 10~MK. The flux-rope plasma appears to consist of a hot core at 10--20~MK and a less hot envelope at 5--10~MK, subject to the confusion of the line-of-sight integration through the optically-thin corona.  Before the seed flux rope appears (the left column in \efig~\ref{fig:dem}), inverse-S shaped loops observed in AIA 94~{\AA} are evident in the 5--10~MK temperature range. The DEM sampled from the sigmoidal flux rope (labeled as `S' in \efig~\ref{fig:dem}) shows that the hot DEM component with $\log T\approx$ 6.8--7.2 increases greatly in magnitude with time, while the cooler components with $\log T<$ 6.8 decrease. In comparison, the DEM sampled from a nearby location off the rope for reference (labeled as `R') changes little with time. Thus, intense heating must be involved in the formation of the flux rope.

That the S-shaped thread continually grows into a bulky sigmoid and later into a CME (\smv~1) is reminiscent of the flux-rope buildup process reported in Gou et al. (2019) \cite{Gou2019}, in which a plasmoid at the upper tip of a pre-existent current sheet embedded in a sheared magnetic arcade continually grows into an oval-shaped bubble and later into a CME. Complementing the side view of the limb eruption on 2013 May 13 in Gou et al. (2019)\cite{Gou2019}, the present on-disk sigmoidal eruption provides a top view of the flux rope formation process. The width of the seed thread is about 2--3$''$ (Fig.~\ref{fig:appear1}), similar to the scale of the leading plasmoid in Gou et al. (2019)\cite{Gou2019}. The temperature structures of the seeds in the two events are also similar, which contain significant hot emissions in the temperature range of $\log T\approx$ 6.7--7.2 (Fig.~\ref{fig:appear2}b and Fig.2 in Gou et al. 2019\cite{Gou2019}). However, the mean temperature of the flux rope in the present event (\efig~\ref{fig:dem}c) could be underestimated due to the contribution from a plethora of overlying loops above it. In addition, owing to the on-disk observation and the unprecedented resolution of IRIS, we observe lower-atmosphere brightenings that evolve synchronously with the development of the sigmoidal flux rope (\efig~\ref{fig:form}). Such brightenings map the footpoints of coronal magnetic field lines undergoing magnetic reconnection, but are seldom observed during the flare precursor phase \cite{WangHM2017}, therefore providing valuable information on the flux-rope buildup process prior to the eruption, which complements the flux-rope buildup process during the eruption \cite{WangWS2017}.

In contrast to two competing pre-eruption magnetic field configurations of CMEs, i.e., sheared arcade vs. flux rope \cite{Patsourakos2020}, the observations that a large-scale flux rope initiates from a seed in both on-disk (e.g., the present study and Wang et al. 2017 \cite{WangWS2017}) and limb (e.g., Gou et al. 2019\cite{Gou2019}) events suggest a hybrid state with a tiny twisted core embedded in a sheared arcade \cite{LiuR2020}. These observations confirm the speculation that there could exist a continuum of intermediate states between the two distinct magnetic configurations \cite{Patsourakos2020}; however, it remains unclear how such an arcade-to-rope transition exactly occurs. The appearance of the seed in the hybrid scenario represents a sudden commencement of the transition from a sheared magentic arcade to a coherent flux rope that starts to build up upon the seed. The buildup process can occur either before (e.g., the event under study) or during the eruption \cite{WangWS2017,Gou2019}. The seed may be a key structure for the the production of highly twisted flux ropes \cite{Priest&Longcope2017} that are detected on the Sun \cite{WangWS2017} or in interplanetary space \cite{WangYM2016}.

\subsection{Partial Eruption}

The central part of the pre-eruption flux rope is aligned with the main PIL of the active region, which is co-spatial with a filament observed in the H$\alpha$ line center by the Big Bear Solar Observatory (BBSO; Fig.~\ref{fig:appear1}d and \efig~\ref{fig:filament}). The filament remains intact beneath the post-flare arcade (\efig~\ref{fig:filament}), which indicates a partial eruption of a double-decker structure in the active region. The lower branch, i.e., the filament, could be either supported by a sheared magnetic arcade or embedded in a flux rope \cite{LiuR2012}. We find a number of candidate BP points at the photospheric PIL that is occupied by the filament (Fig.~\ref{fig:appear1}f). In addition, significant magnetic flux cancellation occurs around the PIL long before the eruption \cite{ChengX2015}, which may produce the BP configuration of magnetic fields \cite{Green2011}. Thus, the BPs may help hold down the filament \cite{Gibson2006}. Both the dextral filament and the inverse S-shaped sigmoid are likely associated with negative magnetic helicity \cite{Martin1998,Rust1999,ChenPF2014}, which also agrees with the accumulation of negative helicity in the active region before the eruption \cite{BiYi2016}.

A question arises as to how and when the double-decker configuration forms. The two branches of the double decker could be adhered to each other in equilibrium before the eruption but split near or during the eruption \cite{Kliem2014}. In this event, we observed that the upper branch of the erupting flux rope originally appears as a shorter thread apparently superposing the low-lying filament (Fig.~\ref{fig:appear1}); its growth associated with footpoint slipping is characterized by the increase of SXR flux and footpoint brightening of chromospheric heating (Fig.~\ref{fig:time} \& \efig~\ref{fig:form}). The footpoint slippage suggests that 3D magnetic reconnection is responsible for the formation of the upper flux rope during the precursor phase before the eruption. Cheng et al. (2015) \cite{ChengX2015} suggests magnetic reconnection occurs in the low solar atmosphere based on the observed Doppler blue shifts. Such reconnections at the bald-patch separatrix surface (BPSS), which consists of magnetic field lines touching the BP points at the photosphere, may lead to the formation of a flux rope atop the structure associated with the filament \cite{Gibson2006,Gibson&Fan2008}.

\subsection{ICME Observation}

The fast full halo CME produced by the sigmoid eruption arrives at the Earth two days after. The Advanced Composition Explorer (ACE) spacecraft observes an interplanetary coronal mass ejection (ICME) structure between Sep 12 22:30~UT and Sep 14 12:00~UT, preceded by a shock front driven by the ICME at 15:30~UT on Sep 12 (\sfig~\ref{fig:icme}). In-situ observations show ICME features \cite{Zurbuchen&Richardson2006} such as enhanced magnetic fields, decreases in proton number density, temperature and plasma $\beta$ (\sfig~\ref{fig:icme}). The bi-directional distribution of suprathermal electrons (\sfig~\ref{fig:icme}h) indicates that the interplanetary structure remains being magnetically connected to the Sun at both ends.

Some features of the ICME also conform to the MC criteria, and it is identified as an MC in a few ICME catalogs \cite{ChiYT2016,Nieves2019,Richardson2010}. However, it does not possess a typical flux-rope configuration. For instance, the magnetic field vectors do not exhibit a large ($\gg30^\circ$) and smooth rotation as typically in MCs \cite{Burlaga1981,Klein&Burlaga1982} (\sfig~\ref{fig:icme}b \& c). Instead, the in-situ observation reveals multiple discontinuities with jumps in both magnetic field and plasma parameters inside the ICME (marked by vertical dotted lines in \sfig~\ref{fig:icme}). Some flux-rope models were used to reconstruct the global magnetic configuration of the ICME. A circular-cylindrical analytical flux-rope model fitting, which was included in a statistical study, obtains a magnetic obstacle with an incomplete rotation less than 90$^\circ$ \cite{Nieves2019}. A torus-shaped flux rope model fitting indicates that the spacecraft may skim the ICME flux rope, covering so small a field rotation that the ICME can be fitted under the assumption of either a left-handed or a right-handed helicity \cite{Marubashi2015,Cho2017}. We also modeled the ICME with two state-of-the-art techniques but did not obtained reasonable results. We first used the velocity-modified cylindrical flux-rope models, adopting either a non-uniform (linear force-free model with Lundquist solution) \citep{WangYM2015} or a uniform (non-linear force-free model) \citep{WangYM2016} twisted magnetic field configuration. Both fittings match poorly with the observations (the normalized root-mean-square difference between the fitting results and observations $\chi_n>0.5$), especially for the magnetic field magnitudes and directions. We also used the Grad-Shafranov (GS) reconstruction technique \cite{HuQ2002}, which does not presume any specific shape of the ICME cross section, but still failed to reconstruct a magnetic flux rope structure. The unsuccessful fittings might be due to limitations of the models, flank crossing of the spacecraft, or complex structures of the ICME itself. 

In this event, the complete replacement of magnetic flux in the preformed flux rope in a drastic fashion makes it hard to imagine that the flux rope can preserve a well-organized structure during its development into the (I)CME. Indeed, the ICME does not possess a typical flux-rope configuration but multiple internal discontinuities (\sfig~\ref{fig:icme}). In-situ detected complex ejecta have been attributed to unfavorable spacecraft trajectories \cite{Zhang2013}, interactions with other ejecta \cite{Burlaga2002} or solar wind \cite{Riley1997}, and the complexity of pre-eruptive structures \cite{Awasthi2018}. Here we suggest that the complex flare reconnections that reshape the flux rope can also contribute to the complexity of the resultant ICME, which agrees with numerical experiments that produce an inherently complex CME \cite{Lynch2008} and may help understand why only about one third of ICMEs possess flux-rope configurations.

\renewcommand{\figurename}{\textbf{Supplementary Fig.}}
\setcounter{figure}{0}
\spacing{1}

\begin{figure}[htbp]
	\centering
	\includegraphics[width=\textwidth]{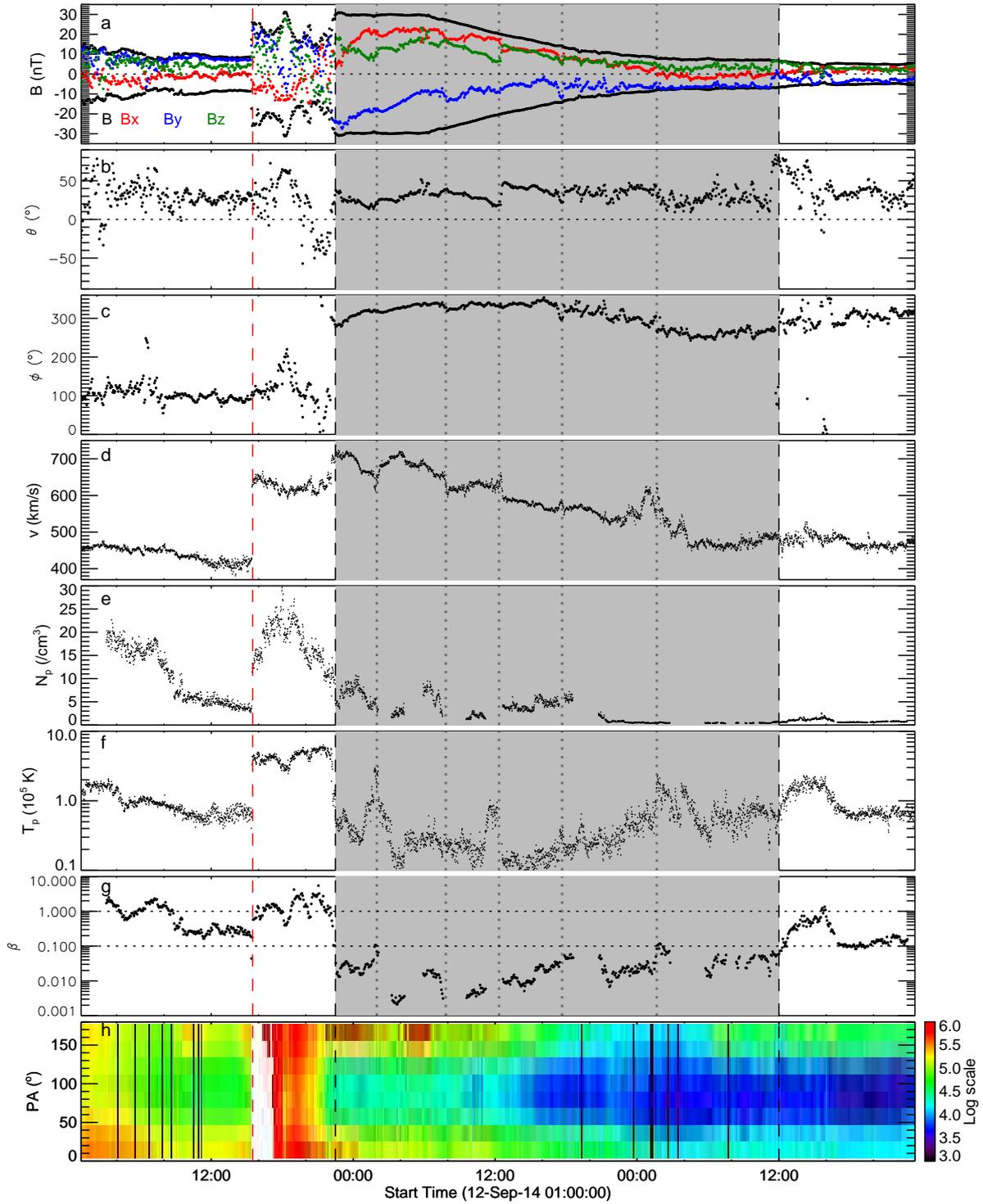}
	\caption{\small \textbf{In situ observations of the ICME.} From the top to bottom are magnetic field magnitude (black) and three rectilinear components (colored), field inclination angle $\theta$ and azimuthal angle $\phi$ in Geocentric Solar Ecliptic (GSE) coordinates, solar wind speed $v$, proton number density $N_p$, proton temperature $T_p$, plasma $\beta$, and pitch angle distribution of $\sim$165~eV suprathermal electrons. In situ data are obtained by ACE spacecraft, but suprathermal electrons by WIND. A shock ahead of the ICME is marked by the red vertical line, and the ICME by the shade. Five vertical dotted lines mark representative discontinuities with jumps in both magnetic field and plasma parameters inside the ICME.}
	\label{fig:icme}
\end{figure}

\clearpage
\spacing{1.25}

\section{Supplementary Videos}

\begin{addendum}
	\item[Supplementary Video 1] SDO/AIA observation of the solar eruption near the disk center. The composite RGB images are made from AIA 131~\AA\ (green), 171~\AA\ (orange), and 1600~\AA\ (blue) channels, which are sensitive to plasma temperatures of about 11, 0.7, and 0.1~MK, respectively.
	\item[Supplementary Video 2] SDO/AIA multi-wavelength observation starting from several hours before the eruption.
	\item[Supplementary Video 3] SDO/AIA and IRIS observation of the flux rope formation and eruption. IRIS's FOV only covers the eastern part of the event, as indicated by the white dotted rectangle. Ratio of AIA 1600 and 1700~\AA\ images are used to highlight the ribbon morphology during the precursor phase before 17:21~UT.
	\item[Supplementary Video 4] SDO/AIA 335~\AA\ and its base ratio images featuring coronal dimmings during the solar eruption.
\end{addendum}


\clearpage

\begin{thebibliography}{10}
	\expandafter\ifx\csname url\endcsname\relax
	\def\url#1{\texttt{#1}}\fi
	\expandafter\ifx\csname urlprefix\endcsname\relax\def\urlprefix{URL }\fi
	\providecommand{\bibinfo}[2]{#2}
	\providecommand{\eprint}[2][]{\url{#2}}
	
	\bibitem{Patsourakos2020}
	\bibinfo{author}{{Patsourakos}, S.} \emph{et~al.}
	\newblock \bibinfo{title}{{Decoding the Pre-Eruptive Magnetic Field
			Configurations of Coronal Mass Ejections}}.
	\newblock \emph{\bibinfo{journal}{\ssr}} \textbf{\bibinfo{volume}{216}},
	\bibinfo{pages}{131} (\bibinfo{year}{2020}).
	
	
	\bibitem{LiuR2020}
	\bibinfo{author}{{Liu}, R.}
	\newblock \bibinfo{title}{{Magnetic flux ropes in the solar corona: structure
			and evolution toward eruption}}.
	\newblock \emph{\bibinfo{journal}{Research in Astronomy and Astrophysics}}
	\textbf{\bibinfo{volume}{20}}, \bibinfo{pages}{165} (\bibinfo{year}{2020}).
	
	
	\bibitem{Priest&Forbes2002}
	\bibinfo{author}{{Priest}, E.~R.} \& \bibinfo{author}{{Forbes}, T.~G.}
	\newblock \bibinfo{title}{{The magnetic nature of solar flares}}.
	\newblock \emph{\bibinfo{journal}{\aapr}} \textbf{\bibinfo{volume}{10}},
	\bibinfo{pages}{313--377} (\bibinfo{year}{2002}).
	
	\bibitem{Temmer2021}
	\bibinfo{author}{{Temmer}, M.}
	\newblock \bibinfo{title}{{Space weather: the solar perspective}}.
	\newblock \emph{\bibinfo{journal}{Living Reviews in Solar Physics}}
	\textbf{\bibinfo{volume}{18}}, \bibinfo{pages}{4} (\bibinfo{year}{2021}).
	
	
	\bibitem{ChengX2017}
	\bibinfo{author}{{Cheng}, X.}, \bibinfo{author}{{Guo}, Y.} \&
	\bibinfo{author}{{Ding}, M.}
	\newblock \bibinfo{title}{{Origin and Structures of Solar Eruptions I: Magnetic
			Flux Rope}}.
	\newblock \emph{\bibinfo{journal}{Science China Earth Sciences}}
	\textbf{\bibinfo{volume}{60}}, \bibinfo{pages}{1383--1407}
	(\bibinfo{year}{2017}).
	
	
	\bibitem{Schmieder2015}
	\bibinfo{author}{{Schmieder}, B.}, \bibinfo{author}{{Aulanier}, G.} \&
	\bibinfo{author}{{Vr{\v{s}}nak}, B.}
	\newblock \bibinfo{title}{{Flare-CME Models: An Observational Perspective
			(Invited Review)}}.
	\newblock \emph{\bibinfo{journal}{\solphys}} \textbf{\bibinfo{volume}{290}},
	\bibinfo{pages}{3457--3486} (\bibinfo{year}{2015}).
	
	\bibitem{Green2018}
	\bibinfo{author}{{Green}, L.~M.}, \bibinfo{author}{{T{\"o}r{\"o}k}, T.},
	\bibinfo{author}{{Vr{\v{s}}nak}, B.}, \bibinfo{author}{{Manchester}, W.} \&
	\bibinfo{author}{{Veronig}, A.}
	\newblock \bibinfo{title}{{The Origin, Early Evolution and Predictability of
			Solar Eruptions}}.
	\newblock \emph{\bibinfo{journal}{\ssr}} \textbf{\bibinfo{volume}{214}},
	\bibinfo{pages}{46} (\bibinfo{year}{2018}).
	
	
	\bibitem{ZhangJ2012}
	\bibinfo{author}{{Zhang}, J.}, \bibinfo{author}{{Cheng}, X.} \&
	\bibinfo{author}{{Ding}, M.-D.}
	\newblock \bibinfo{title}{{Observation of an evolving magnetic flux rope before
			and during a solar eruption}}.
	\newblock \emph{\bibinfo{journal}{Nature Communications}}
	\textbf{\bibinfo{volume}{3}}, \bibinfo{pages}{747} (\bibinfo{year}{2012}).
	
	
	\bibitem{Patsourakos2013}
	\bibinfo{author}{{Patsourakos}, S.}, \bibinfo{author}{{Vourlidas}, A.} \&
	\bibinfo{author}{{Stenborg}, G.}
	\newblock \bibinfo{title}{{Direct Evidence for a Fast Coronal Mass Ejection
			Driven by the Prior Formation and Subsequent Destabilization of a Magnetic
			Flux Rope}}.
	\newblock \emph{\bibinfo{journal}{\apj}} \textbf{\bibinfo{volume}{764}},
	\bibinfo{pages}{125} (\bibinfo{year}{2013}).
	
	
	\bibitem{Moore2001}
	\bibinfo{author}{{Moore}, R.~L.}, \bibinfo{author}{{Sterling}, A.~C.},
	\bibinfo{author}{{Hudson}, H.~S.} \& \bibinfo{author}{{Lemen}, J.~R.}
	\newblock \bibinfo{title}{{Onset of the Magnetic Explosion in Solar Flares and
			Coronal Mass Ejections}}.
	\newblock \emph{\bibinfo{journal}{\apj}} \textbf{\bibinfo{volume}{552}},
	\bibinfo{pages}{833--848} (\bibinfo{year}{2001}).
	
	\bibitem{Karpen2012}
	\bibinfo{author}{{Karpen}, J.~T.}, \bibinfo{author}{{Antiochos}, S.~K.} \&
	\bibinfo{author}{{DeVore}, C.~R.}
	\newblock \bibinfo{title}{{The Mechanisms for the Onset and Explosive Eruption
			of Coronal Mass Ejections and Eruptive Flares}}.
	\newblock \emph{\bibinfo{journal}{\apj}} \textbf{\bibinfo{volume}{760}},
	\bibinfo{pages}{81} (\bibinfo{year}{2012}).
	
	\bibitem{Gou2019}
	\bibinfo{author}{{Gou}, T.}, \bibinfo{author}{{Liu}, R.},
	\bibinfo{author}{{Kliem}, B.}, \bibinfo{author}{{Wang}, Y.} \&
	\bibinfo{author}{{Veronig}, A.~M.}
	\newblock \bibinfo{title}{{The Birth of A Coronal Mass Ejection}}.
	\newblock \emph{\bibinfo{journal}{Science Advances}}
	\textbf{\bibinfo{volume}{5}}, \bibinfo{pages}{7004} (\bibinfo{year}{2019}).
	
	
	\bibitem{Kopp&Pneuman1976}
	\bibinfo{author}{{Kopp}, R.~A.} \& \bibinfo{author}{{Pneuman}, G.~W.}
	\newblock \bibinfo{title}{{Magnetic reconnection in the corona and the loop
			prominence phenomenon.}}
	\newblock \emph{\bibinfo{journal}{\solphys}} \textbf{\bibinfo{volume}{50}},
	\bibinfo{pages}{85--98} (\bibinfo{year}{1976}).
	
	\bibitem{Lin&Forbes2000}
	\bibinfo{author}{{Lin}, J.} \& \bibinfo{author}{{Forbes}, T.~G.}
	\newblock \bibinfo{title}{{Effects of reconnection on the coronal mass ejection
			process}}.
	\newblock \emph{\bibinfo{journal}{\jgr}} \textbf{\bibinfo{volume}{105}},
	\bibinfo{pages}{2375--2392} (\bibinfo{year}{2000}).
	
	\bibitem{Demoulin2006}
	\bibinfo{author}{{D{\'e}moulin}, P.}
	\newblock \bibinfo{title}{{Extending the concept of separatrices to QSLs for
			magnetic reconnection}}.
	\newblock \emph{\bibinfo{journal}{Advances in Space Research}}
	\textbf{\bibinfo{volume}{37}}, \bibinfo{pages}{1269--1282}
	(\bibinfo{year}{2006}).
	
	\bibitem{Aulanier2012}
	\bibinfo{author}{{Aulanier}, G.}, \bibinfo{author}{{Janvier}, M.} \&
	\bibinfo{author}{{Schmieder}, B.}
	\newblock \bibinfo{title}{{The standard flare model in three dimensions. I.
			Strong-to-weak shear transition in post-flare loops}}.
	\newblock \emph{\bibinfo{journal}{\aap}} \textbf{\bibinfo{volume}{543}},
	\bibinfo{pages}{A110} (\bibinfo{year}{2012}).
	
	\bibitem{Aulanier2013}
	\bibinfo{author}{{Aulanier}, G.} \emph{et~al.}
	\newblock \bibinfo{title}{{The standard flare model in three dimensions. II.
			Upper limit on solar flare energy}}.
	\newblock \emph{\bibinfo{journal}{\aap}} \textbf{\bibinfo{volume}{549}},
	\bibinfo{pages}{A66} (\bibinfo{year}{2013}).
	
	
	\bibitem{Janvier2013}
	\bibinfo{author}{{Janvier}, M.}, \bibinfo{author}{{Aulanier}, G.},
	\bibinfo{author}{{Pariat}, E.} \& \bibinfo{author}{{D{\'e}moulin}, P.}
	\newblock \bibinfo{title}{{The standard flare model in three dimensions. III.
			Slip-running reconnection properties}}.
	\newblock \emph{\bibinfo{journal}{\aap}} \textbf{\bibinfo{volume}{555}},
	\bibinfo{pages}{A77} (\bibinfo{year}{2013}).
	
	
	\bibitem{Titov2002}
	\bibinfo{author}{{Titov}, V.~S.}, \bibinfo{author}{{Hornig}, G.} \&
	\bibinfo{author}{{D{\'e}moulin}, P.}
	\newblock \bibinfo{title}{{Theory of magnetic connectivity in the solar
			corona}}.
	\newblock \emph{\bibinfo{journal}{Journal of Geophysical Research (Space
			Physics)}} \textbf{\bibinfo{volume}{107}}, \bibinfo{pages}{1164}
	(\bibinfo{year}{2002}).
	
	\bibitem{Janvier2014}
	\bibinfo{author}{{Janvier}, M.} \emph{et~al.}
	\newblock \bibinfo{title}{{Electric Currents in Flare Ribbons: Observations and
			Three-dimensional Standard Model}}.
	\newblock \emph{\bibinfo{journal}{\apj}} \textbf{\bibinfo{volume}{788}},
	\bibinfo{pages}{60} (\bibinfo{year}{2014}).
	
	
	\bibitem{WangWS2017}
	\bibinfo{author}{{Wang}, W.} \emph{et~al.}
	\newblock \bibinfo{title}{{Buildup of a highly twisted magnetic flux rope
			during a solar eruption}}.
	\newblock \emph{\bibinfo{journal}{Nature Communications}}
	\textbf{\bibinfo{volume}{8}}, \bibinfo{pages}{1330} (\bibinfo{year}{2017}).
	
	\bibitem{Veronig2019}
	\bibinfo{author}{{Veronig}, A.~M.}, \bibinfo{author}{{G{\"o}m{\"o}ry}, P.},
	\bibinfo{author}{{Dissauer}, K.}, \bibinfo{author}{{Temmer}, M.} \&
	\bibinfo{author}{{Vanninathan}, K.}
	\newblock \bibinfo{title}{{Spectroscopy and Differential Emission Measure
			Diagnostics of a Coronal Dimming Associated with a Fast Halo CME}}.
	\newblock \emph{\bibinfo{journal}{\apj}} \textbf{\bibinfo{volume}{879}},
	\bibinfo{pages}{85} (\bibinfo{year}{2019}).
	
	
	\bibitem{XingC2020}
	\bibinfo{author}{{Xing}, C.}, \bibinfo{author}{{Cheng}, X.} \&
	\bibinfo{author}{{Ding}, M.~D.}
	\newblock \bibinfo{title}{{Evolution of the Toroidal Flux of CME Flux Ropes
			during Eruption}}.
	\newblock \emph{\bibinfo{journal}{The Innovation}}
	\textbf{\bibinfo{volume}{1}}, \bibinfo{pages}{100059} (\bibinfo{year}{2020}).
	
	
	\bibitem{Burlaga1981}
	\bibinfo{author}{{Burlaga}, L.}, \bibinfo{author}{{Sittler}, E.},
	\bibinfo{author}{{Mariani}, F.} \& \bibinfo{author}{{Schwenn}, R.}
	\newblock \bibinfo{title}{{Magnetic loop behind an interplanetary shock:
			Voyager, Helios, and IMP 8 observations}}.
	\newblock \emph{\bibinfo{journal}{\jgr}} \textbf{\bibinfo{volume}{86}},
	\bibinfo{pages}{6673--6684} (\bibinfo{year}{1981}).
	
	\bibitem{Klein&Burlaga1982}
	\bibinfo{author}{{Klein}, L.~W.} \& \bibinfo{author}{{Burlaga}, L.~F.}
	\newblock \bibinfo{title}{{Interplanetary magnetic clouds at 1 AU}}.
	\newblock \emph{\bibinfo{journal}{\jgr}} \textbf{\bibinfo{volume}{87}},
	\bibinfo{pages}{613--624} (\bibinfo{year}{1982}).
	
	\bibitem{ChiYT2016}
	\bibinfo{author}{{Chi}, Y.} \emph{et~al.}
	\newblock \bibinfo{title}{{Statistical Study of the Interplanetary Coronal Mass
			Ejections from 1995 to 2015}}.
	\newblock \emph{\bibinfo{journal}{\solphys}} \textbf{\bibinfo{volume}{291}},
	\bibinfo{pages}{2419--2439} (\bibinfo{year}{2016}).
	
	
	\bibitem{Vourlidas2013}
	\bibinfo{author}{{Vourlidas}, A.}, \bibinfo{author}{{Lynch}, B.~J.},
	\bibinfo{author}{{Howard}, R.~A.} \& \bibinfo{author}{{Li}, Y.}
	\newblock \bibinfo{title}{{How Many CMEs Have Flux Ropes? Deciphering the
			Signatures of Shocks, Flux Ropes, and Prominences in Coronagraph Observations
			of CMEs}}.
	\newblock \emph{\bibinfo{journal}{\solphys}} \textbf{\bibinfo{volume}{284}},
	\bibinfo{pages}{179--201} (\bibinfo{year}{2013}).
	
	
	\bibitem{SongHQ2020}
	\bibinfo{author}{{Song}, H.~Q.} \emph{et~al.}
	\newblock \bibinfo{title}{{Do All Interplanetary Coronal Mass Ejections Have a
			Magnetic Flux Rope Structure Near 1 au?}}
	\newblock \emph{\bibinfo{journal}{\apjl}} \textbf{\bibinfo{volume}{901}},
	\bibinfo{pages}{L21} (\bibinfo{year}{2020}).
	
	
	\bibitem{ChengX2015}
	\bibinfo{author}{{Cheng}, X.}, \bibinfo{author}{{Ding}, M.~D.} \&
	\bibinfo{author}{{Fang}, C.}
	\newblock \bibinfo{title}{{Imaging and Spectroscopic Diagnostics on the
			Formation of Two Magnetic Flux Ropes Revealed by SDO/AIA and IRIS}}.
	\newblock \emph{\bibinfo{journal}{\apj}} \textbf{\bibinfo{volume}{804}},
	\bibinfo{pages}{82} (\bibinfo{year}{2015}).
	
	
	\bibitem{Dudik2016}
	\bibinfo{author}{{Dud{\'\i}k}, J.} \emph{et~al.}
	\newblock \bibinfo{title}{{Slipping Magnetic Reconnection, Chromospheric
			Evaporation, Implosion, and Precursors in the 2014 September 10 X1.6-Class
			Solar Flare}}.
	\newblock \emph{\bibinfo{journal}{\apj}} \textbf{\bibinfo{volume}{823}},
	\bibinfo{pages}{41} (\bibinfo{year}{2016}).
	
	
	\bibitem{ZhaoJ2016}
	\bibinfo{author}{{Zhao}, J.} \emph{et~al.}
	\newblock \bibinfo{title}{{Hooked Flare Ribbons and Flux-rope-related QSL
			Footprints}}.
	\newblock \emph{\bibinfo{journal}{\apj}} \textbf{\bibinfo{volume}{823}},
	\bibinfo{pages}{62} (\bibinfo{year}{2016}).
	
	
	\bibitem{Demoulin1996fluxtube}
	\bibinfo{author}{{D{\'e}moulin}, P.}, \bibinfo{author}{{Priest}, E.~R.} \&
	\bibinfo{author}{{Lonie}, D.~P.}
	\newblock \bibinfo{title}{{Three-dimensional magnetic reconnection without null
			points 2. Application to twisted flux tubes}}.
	\newblock \emph{\bibinfo{journal}{\jgr}} \textbf{\bibinfo{volume}{101}},
	\bibinfo{pages}{7631--7646} (\bibinfo{year}{1996}).
	
	\bibitem{LiT2015}
	\bibinfo{author}{{Li}, T.} \& \bibinfo{author}{{Zhang}, J.}
	\newblock \bibinfo{title}{{Quasi-periodic Slipping Magnetic Reconnection During
			an X-class Solar Flare Observed by the Solar Dynamics Observatory and
			Interface Region Imaging Spectrograph}}.
	\newblock \emph{\bibinfo{journal}{\apjl}} \textbf{\bibinfo{volume}{804}},
	\bibinfo{pages}{L8} (\bibinfo{year}{2015}).
	
	
	\bibitem{Aulanier&Dudik2019}
	\bibinfo{author}{{Aulanier}, G.} \& \bibinfo{author}{{Dud{\'\i}k}, J.}
	\newblock \bibinfo{title}{{Drifting of the line-tied footpoints of CME
			flux-ropes}}.
	\newblock \emph{\bibinfo{journal}{\aap}} \textbf{\bibinfo{volume}{621}},
	\bibinfo{pages}{A72} (\bibinfo{year}{2019}).
	
	
	\bibitem{Dudik2019}
	\bibinfo{author}{{Dud{\'\i}k}, J.}, \bibinfo{author}{{L{\"o}rin{\v{c}}{\'\i}k},
		J.}, \bibinfo{author}{{Aulanier}, G.}, \bibinfo{author}{{Zemanov{\'a}}, A.}
	\& \bibinfo{author}{{Schmieder}, B.}
	\newblock \bibinfo{title}{{Observation of All Pre- and Post-reconnection
			Structures Involved in Three-dimensional Reconnection Geometries in Solar
			Eruptions}}.
	\newblock \emph{\bibinfo{journal}{\apj}} \textbf{\bibinfo{volume}{887}},
	\bibinfo{pages}{71} (\bibinfo{year}{2019}).
	
	
	\bibitem{vanDriel2014}
	\bibinfo{author}{{van Driel-Gesztelyi}, L.} \emph{et~al.}
	\newblock \bibinfo{title}{{Coronal Magnetic Reconnection Driven by CME
			Expansion{\textemdash}the 2011 June 7 Event}}.
	\newblock \emph{\bibinfo{journal}{\apj}} \textbf{\bibinfo{volume}{788}},
	\bibinfo{pages}{85} (\bibinfo{year}{2014}).
	
	
	\bibitem{Torok2018}
	\bibinfo{author}{{T{\"o}r{\"o}k}, T.} \emph{et~al.}
	\newblock \bibinfo{title}{{Sun-to-Earth MHD Simulation of the 2000 July 14
			{\textquotedblleft}Bastille Day{\textquotedblright} Eruption}}.
	\newblock \emph{\bibinfo{journal}{\apj}} \textbf{\bibinfo{volume}{856}},
	\bibinfo{pages}{75} (\bibinfo{year}{2018}).
	
	
	\bibitem{QiuJ2007}
	\bibinfo{author}{{Qiu}, J.}, \bibinfo{author}{{Hu}, Q.},
	\bibinfo{author}{{Howard}, T.~A.} \& \bibinfo{author}{{Yurchyshyn}, V.~B.}
	\newblock \bibinfo{title}{{On the Magnetic Flux Budget in Low-Corona Magnetic
			Reconnection and Interplanetary Coronal Mass Ejections}}.
	\newblock \emph{\bibinfo{journal}{\apj}} \textbf{\bibinfo{volume}{659}},
	\bibinfo{pages}{758--772} (\bibinfo{year}{2007}).
	
	\bibitem{WangYM2015}
	\bibinfo{author}{{Wang}, Y.}, \bibinfo{author}{{Zhou}, Z.},
	\bibinfo{author}{{Shen}, C.}, \bibinfo{author}{{Liu}, R.} \&
	\bibinfo{author}{{Wang}, S.}
	\newblock \bibinfo{title}{{Investigating plasma motion of magnetic clouds at 1
			AU through a velocity-modified cylindrical force-free flux rope model}}.
	\newblock \emph{\bibinfo{journal}{Journal of Geophysical Research (Space
			Physics)}} \textbf{\bibinfo{volume}{120}}, \bibinfo{pages}{1543--1565}
	(\bibinfo{year}{2015}).
	
	
	\bibitem{Kazachenko2017}
	\bibinfo{author}{{Kazachenko}, M.~D.}, \bibinfo{author}{{Lynch}, B.~J.},
	\bibinfo{author}{{Welsch}, B.~T.} \& \bibinfo{author}{{Sun}, X.}
	\newblock \bibinfo{title}{{A Database of Flare Ribbon Properties from the Solar
			Dynamics Observatory. I. Reconnection Flux}}.
	\newblock \emph{\bibinfo{journal}{\apj}} \textbf{\bibinfo{volume}{845}},
	\bibinfo{pages}{49} (\bibinfo{year}{2017}).
	
	
	\bibitem{Tschernitz2018}
	\bibinfo{author}{{Tschernitz}, J.}, \bibinfo{author}{{Veronig}, A.~M.},
	\bibinfo{author}{{Thalmann}, J.~K.}, \bibinfo{author}{{Hinterreiter}, J.} \&
	\bibinfo{author}{{P{\"o}tzi}, W.}
	\newblock \bibinfo{title}{{Reconnection Fluxes in Eruptive and Confined Flares
			and Implications for Superflares on the Sun}}.
	\newblock \emph{\bibinfo{journal}{\apj}} \textbf{\bibinfo{volume}{853}},
	\bibinfo{pages}{41} (\bibinfo{year}{2018}).
	
	
	\bibitem{Jiang2021}
	\bibinfo{author}{{Jiang}, C.} \emph{et~al.}
	\newblock \bibinfo{title}{{A fundamental mechanism of solar eruption
			initiation}}.
	\newblock \emph{\bibinfo{journal}{Nature Astronomy}}
	\textbf{\bibinfo{volume}{5}}, \bibinfo{pages}{1126--1138}
	(\bibinfo{year}{2021}).
	
	
	\bibitem{Lemen2012}
	\bibinfo{author}{{Lemen}, J.~R.} \emph{et~al.}
	\newblock \bibinfo{title}{{The Atmospheric Imaging Assembly (AIA) on the Solar
			Dynamics Observatory (SDO)}}.
	\newblock \emph{\bibinfo{journal}{\solphys}} \textbf{\bibinfo{volume}{275}},
	\bibinfo{pages}{17--40} (\bibinfo{year}{2012}).
	
	\bibitem{Pesnell2012}
	\bibinfo{author}{{Pesnell}, W.~D.}, \bibinfo{author}{{Thompson}, B.~J.} \&
	\bibinfo{author}{{Chamberlin}, P.~C.}
	\newblock \bibinfo{title}{{The Solar Dynamics Observatory (SDO)}}.
	\newblock \emph{\bibinfo{journal}{\solphys}} \textbf{\bibinfo{volume}{275}},
	\bibinfo{pages}{3--15} (\bibinfo{year}{2012}).
	
	\bibitem{Schou2012}
	\bibinfo{author}{{Schou}, J.} \emph{et~al.}
	\newblock \bibinfo{title}{{Design and Ground Calibration of the Helioseismic
			and Magnetic Imager (HMI) Instrument on the Solar Dynamics Observatory
			(SDO)}}.
	\newblock \emph{\bibinfo{journal}{\solphys}} \textbf{\bibinfo{volume}{275}},
	\bibinfo{pages}{229--259} (\bibinfo{year}{2012}).
	
	\bibitem{Titov1993}
	\bibinfo{author}{{Titov}, V.~S.}, \bibinfo{author}{{Priest}, E.~R.} \&
	\bibinfo{author}{{Demoulin}, P.}
	\newblock \bibinfo{title}{{Conditions for the appearance of ``bald patches'' at
			the solar surface}}.
	\newblock \emph{\bibinfo{journal}{\aap}} \textbf{\bibinfo{volume}{276}},
	\bibinfo{pages}{564} (\bibinfo{year}{1993}).
	
	\bibitem{Titov&Demoulin1999}
	\bibinfo{author}{{Titov}, V.~S.} \& \bibinfo{author}{{D{\'e}moulin}, P.}
	\newblock \bibinfo{title}{{Basic topology of twisted magnetic configurations in
			solar flares}}.
	\newblock \emph{\bibinfo{journal}{\aap}} \textbf{\bibinfo{volume}{351}},
	\bibinfo{pages}{707--720} (\bibinfo{year}{1999}).
	
	\bibitem{DePontieu2014}
	\bibinfo{author}{{De Pontieu}, B.} \emph{et~al.}
	\newblock \bibinfo{title}{{The Interface Region Imaging Spectrograph (IRIS)}}.
	\newblock \emph{\bibinfo{journal}{\solphys}} \textbf{\bibinfo{volume}{289}},
	\bibinfo{pages}{2733--2779} (\bibinfo{year}{2014}).
	
	
	\bibitem{Cheung2015}
	\bibinfo{author}{{Cheung}, M. C.~M.} \emph{et~al.}
	\newblock \bibinfo{title}{{Thermal Diagnostics with the Atmospheric Imaging
			Assembly on board the Solar Dynamics Observatory: A Validated Method for
			Differential Emission Measure Inversions}}.
	\newblock \emph{\bibinfo{journal}{\apj}} \textbf{\bibinfo{volume}{807}},
	\bibinfo{pages}{143} (\bibinfo{year}{2015}).
	
	
	\bibitem{SuY2018}
	\bibinfo{author}{{Su}, Y.} \emph{et~al.}
	\newblock \bibinfo{title}{{Determination of Differential Emission Measure from
			Solar Extreme Ultraviolet Images}}.
	\newblock \emph{\bibinfo{journal}{\apjl}} \textbf{\bibinfo{volume}{856}},
	\bibinfo{pages}{L17} (\bibinfo{year}{2018}).
	
	\bibitem{Gou2015}
	\bibinfo{author}{{Gou}, T.}, \bibinfo{author}{{Liu}, R.} \&
	\bibinfo{author}{{Wang}, Y.}
	\newblock \bibinfo{title}{{Do All Candle-Flame-Shaped Flares Have the Same
			Temperature Distribution?}}
	\newblock \emph{\bibinfo{journal}{\solphys}} \textbf{\bibinfo{volume}{290}},
	\bibinfo{pages}{2211--2230} (\bibinfo{year}{2015}).
	
\end{thebibliography}

\clearpage
\small


\begin{thebibliography}{10}
	\expandafter\ifx\csname url\endcsname\relax
	\def\url#1{\texttt{#1}}\fi
	\expandafter\ifx\csname urlprefix\endcsname\relax\def\urlprefix{URL }\fi
	\providecommand{\bibinfo}[2]{#2}
	\providecommand{\eprint}[2][]{\url{#2}}
	
	\makeatletter
	\addtocounter{NAT@ctr}{51}
	\makeatother
	
	\bibitem{Demoulin1996fluxtube}
	\bibinfo{author}{{D{\'e}moulin}, P.}, \bibinfo{author}{{Priest}, E.~R.} \&
	\bibinfo{author}{{Lonie}, D.~P.}
	\newblock \bibinfo{title}{{Three-dimensional magnetic reconnection without null
			points 2. Application to twisted flux tubes}}.
	\newblock \emph{\bibinfo{journal}{\jgr}} \textbf{\bibinfo{volume}{101}},
	\bibinfo{pages}{7631--7646} (\bibinfo{year}{1996}).
	
	\bibitem{WangWS2017}
	\bibinfo{author}{{Wang}, W.} \emph{et~al.}
	\newblock \bibinfo{title}{{Buildup of a highly twisted magnetic flux rope
			during a solar eruption}}.
	\newblock \emph{\bibinfo{journal}{Nature Communications}}
	\textbf{\bibinfo{volume}{8}}, \bibinfo{pages}{1330} (\bibinfo{year}{2017}).
	
	\bibitem{vanBallegooijen1989}
	\bibinfo{author}{{van Ballegooijen}, A.~A.} \& \bibinfo{author}{{Martens},
		P.~C.~H.}
	\newblock \bibinfo{title}{{Formation and Eruption of Solar Prominences}}.
	\newblock \emph{\bibinfo{journal}{\apj}} \textbf{\bibinfo{volume}{343}},
	\bibinfo{pages}{971} (\bibinfo{year}{1989}).
	
	\bibitem{Moore2001}
	\bibinfo{author}{{Moore}, R.~L.}, \bibinfo{author}{{Sterling}, A.~C.},
	\bibinfo{author}{{Hudson}, H.~S.} \& \bibinfo{author}{{Lemen}, J.~R.}
	\newblock \bibinfo{title}{{Onset of the Magnetic Explosion in Solar Flares and
			Coronal Mass Ejections}}.
	\newblock \emph{\bibinfo{journal}{\apj}} \textbf{\bibinfo{volume}{552}},
	\bibinfo{pages}{833--848} (\bibinfo{year}{2001}).
	
	\bibitem{Green2011}
	\bibinfo{author}{{Green}, L.~M.}, \bibinfo{author}{{Kliem}, B.} \&
	\bibinfo{author}{{Wallace}, A.~J.}
	\newblock \bibinfo{title}{{Photospheric flux cancellation and associated flux
			rope formation and eruption}}.
	\newblock \emph{\bibinfo{journal}{\aap}} \textbf{\bibinfo{volume}{526}},
	\bibinfo{pages}{A2} (\bibinfo{year}{2011}).
	
	\bibitem{ChengX2015}
	\bibinfo{author}{{Cheng}, X.}, \bibinfo{author}{{Ding}, M.~D.} \&
	\bibinfo{author}{{Fang}, C.}
	\newblock \bibinfo{title}{{Imaging and Spectroscopic Diagnostics on the
			Formation of Two Magnetic Flux Ropes Revealed by SDO/AIA and IRIS}}.
	\newblock \emph{\bibinfo{journal}{\apj}} \textbf{\bibinfo{volume}{804}},
	\bibinfo{pages}{82} (\bibinfo{year}{2015}).
	
	\bibitem{Aulanier2007}
	\bibinfo{author}{{Aulanier}, G.} \emph{et~al.}
	\newblock \bibinfo{title}{{Slipping Magnetic Reconnection in Coronal Loops}}.
	\newblock \emph{\bibinfo{journal}{Science}} \textbf{\bibinfo{volume}{318}},
	\bibinfo{pages}{1588} (\bibinfo{year}{2007}).
	
	\bibitem{Gou2019}
	\bibinfo{author}{{Gou}, T.}, \bibinfo{author}{{Liu}, R.},
	\bibinfo{author}{{Kliem}, B.}, \bibinfo{author}{{Wang}, Y.} \&
	\bibinfo{author}{{Veronig}, A.~M.}
	\newblock \bibinfo{title}{{The Birth of A Coronal Mass Ejection}}.
	\newblock \emph{\bibinfo{journal}{Science Advances}}
	\textbf{\bibinfo{volume}{5}}, \bibinfo{pages}{7004} (\bibinfo{year}{2019}).
	
	\bibitem{WangHM2017}
	\bibinfo{author}{{Wang}, H.} \emph{et~al.}
	\newblock \bibinfo{title}{{High-resolution observations of flare precursors in
			the low solar atmosphere}}.
	\newblock \emph{\bibinfo{journal}{Nature Astronomy}}
	\textbf{\bibinfo{volume}{1}}, \bibinfo{pages}{0085} (\bibinfo{year}{2017}).
	
	\bibitem{Patsourakos2020}
	\bibinfo{author}{{Patsourakos}, S.} \emph{et~al.}
	\newblock \bibinfo{title}{{Decoding the Pre-Eruptive Magnetic Field
			Configurations of Coronal Mass Ejections}}.
	\newblock \emph{\bibinfo{journal}{\ssr}} \textbf{\bibinfo{volume}{216}},
	\bibinfo{pages}{131} (\bibinfo{year}{2020}).
	
	\bibitem{LiuR2020}
	\bibinfo{author}{{Liu}, R.}
	\newblock \bibinfo{title}{{Magnetic flux ropes in the solar corona: structure
			and evolution toward eruption}}.
	\newblock \emph{\bibinfo{journal}{Research in Astronomy and Astrophysics}}
	\textbf{\bibinfo{volume}{20}}, \bibinfo{pages}{165} (\bibinfo{year}{2020}).
	
	\bibitem{Priest&Longcope2017}
	\bibinfo{author}{{Priest}, E.~R.} \& \bibinfo{author}{{Longcope}, D.~W.}
	\newblock \bibinfo{title}{{Flux-Rope Twist in Eruptive Flares and CMEs: Due to
			Zipper and Main-Phase Reconnection}}.
	\newblock \emph{\bibinfo{journal}{\solphys}} \textbf{\bibinfo{volume}{292}},
	\bibinfo{pages}{25} (\bibinfo{year}{2017}).
	
	\bibitem{WangYM2016}
	\bibinfo{author}{{Wang}, Y.} \emph{et~al.}
	\newblock \bibinfo{title}{{On the twists of interplanetary magnetic flux ropes
			observed at 1 AU}}.
	\newblock \emph{\bibinfo{journal}{Journal of Geophysical Research (Space
			Physics)}} \textbf{\bibinfo{volume}{121}}, \bibinfo{pages}{9316--9339}
	(\bibinfo{year}{2016}).
	
	\bibitem{LiuR2012}
	\bibinfo{author}{{Liu}, R.} \emph{et~al.}
	\newblock \bibinfo{title}{{Slow Rise and Partial Eruption of a Double-decker
			Filament. I. Observations and Interpretation}}.
	\newblock \emph{\bibinfo{journal}{\apj}} \textbf{\bibinfo{volume}{756}},
	\bibinfo{pages}{59} (\bibinfo{year}{2012}).
	
	\bibitem{Gibson2006}
	\bibinfo{author}{{Gibson}, S.~E.}, \bibinfo{author}{{Fan}, Y.},
	\bibinfo{author}{{T{\"o}r{\"o}k}, T.} \& \bibinfo{author}{{Kliem}, B.}
	\newblock \bibinfo{title}{{The Evolving Sigmoid: Evidence for Magnetic Flux
			Ropes in the Corona Before, During, and After CMES}}.
	\newblock \emph{\bibinfo{journal}{\ssr}} \textbf{\bibinfo{volume}{124}},
	\bibinfo{pages}{131--144} (\bibinfo{year}{2006}).
	
	\bibitem{Martin1998}
	\bibinfo{author}{{Martin}, S.~F.}
	\newblock \bibinfo{title}{{Conditions for the Formation and Maintenance of
			Filaments (Invited Review)}}.
	\newblock \emph{\bibinfo{journal}{\solphys}} \textbf{\bibinfo{volume}{182}},
	\bibinfo{pages}{107--137} (\bibinfo{year}{1998}).
	
	\bibitem{Rust1999}
	\bibinfo{author}{{Rust}, D.~M.}
	\newblock \bibinfo{title}{{Magnetic Helicity in Solar Filaments and Coronal
			Mass Ejections}}.
	\newblock \emph{\bibinfo{journal}{Washington DC American Geophysical Union
			Geophysical Monograph Series}} \textbf{\bibinfo{volume}{111}},
	\bibinfo{pages}{221} (\bibinfo{year}{1999}).
	
	\bibitem{ChenPF2014}
	\bibinfo{author}{{Chen}, P.~F.}, \bibinfo{author}{{Harra}, L.~K.} \&
	\bibinfo{author}{{Fang}, C.}
	\newblock \bibinfo{title}{{Imaging and Spectroscopic Observations of a Filament
			Channel and the Implications for the Nature of Counter-streamings}}.
	\newblock \emph{\bibinfo{journal}{\apj}} \textbf{\bibinfo{volume}{784}},
	\bibinfo{pages}{50} (\bibinfo{year}{2014}).
	
	\bibitem{BiYi2016}
	\bibinfo{author}{{Bi}, Y.} \emph{et~al.}
	\newblock \bibinfo{title}{{Observation of a reversal of rotation in a sunspot
			during a solar flare}}.
	\newblock \emph{\bibinfo{journal}{Nature Communications}}
	\textbf{\bibinfo{volume}{7}}, \bibinfo{pages}{13798} (\bibinfo{year}{2016}).
	
	\bibitem{Kliem2014}
	\bibinfo{author}{{Kliem}, B.} \emph{et~al.}
	\newblock \bibinfo{title}{{Slow Rise and Partial Eruption of a Double-decker
			Filament. II. A Double Flux Rope Model}}.
	\newblock \emph{\bibinfo{journal}{\apj}} \textbf{\bibinfo{volume}{792}},
	\bibinfo{pages}{107} (\bibinfo{year}{2014}).
	
	\bibitem{Gibson&Fan2008}
	\bibinfo{author}{{Gibson}, S.~E.} \& \bibinfo{author}{{Fan}, Y.}
	\newblock \bibinfo{title}{{Partially ejected flux ropes: Implications for
			interplanetary coronal mass ejections}}.
	\newblock \emph{\bibinfo{journal}{Journal of Geophysical Research (Space
			Physics)}} \textbf{\bibinfo{volume}{113}}, \bibinfo{pages}{A09103}
	(\bibinfo{year}{2008}).
	
	\bibitem{Zurbuchen&Richardson2006}
	\bibinfo{author}{{Zurbuchen}, T.~H.} \& \bibinfo{author}{{Richardson}, I.~G.}
	\newblock \bibinfo{title}{{In-Situ Solar Wind and Magnetic Field Signatures of
			Interplanetary Coronal Mass Ejections}}.
	\newblock \emph{\bibinfo{journal}{\ssr}} \textbf{\bibinfo{volume}{123}},
	\bibinfo{pages}{31--43} (\bibinfo{year}{2006}).
	
	\bibitem{ChiYT2016}
	\bibinfo{author}{{Chi}, Y.} \emph{et~al.}
	\newblock \bibinfo{title}{{Statistical Study of the Interplanetary Coronal Mass
			Ejections from 1995 to 2015}}.
	\newblock \emph{\bibinfo{journal}{\solphys}} \textbf{\bibinfo{volume}{291}},
	\bibinfo{pages}{2419--2439} (\bibinfo{year}{2016}).
	
	\bibitem{Nieves2019}
	\bibinfo{author}{{Nieves-Chinchilla}, T.} \emph{et~al.}
	\newblock \bibinfo{title}{{Unraveling the Internal Magnetic Field Structure of
			the Earth-directed Interplanetary Coronal Mass Ejections During 1995 -
			2015}}.
	\newblock \emph{\bibinfo{journal}{\solphys}} \textbf{\bibinfo{volume}{294}},
	\bibinfo{pages}{89} (\bibinfo{year}{2019}).
	
	\bibitem{Richardson2010}
	\bibinfo{author}{{Richardson}, I.~G.} \& \bibinfo{author}{{Cane}, H.~V.}
	\newblock \bibinfo{title}{{Near-Earth Interplanetary Coronal Mass Ejections
			During Solar Cycle 23 (1996 - 2009): Catalog and Summary of Properties}}.
	\newblock \emph{\bibinfo{journal}{\solphys}} \textbf{\bibinfo{volume}{264}},
	\bibinfo{pages}{189--237} (\bibinfo{year}{2010}).
	
	\bibitem{Burlaga1981}
	\bibinfo{author}{{Burlaga}, L.}, \bibinfo{author}{{Sittler}, E.},
	\bibinfo{author}{{Mariani}, F.} \& \bibinfo{author}{{Schwenn}, R.}
	\newblock \bibinfo{title}{{Magnetic loop behind an interplanetary shock:
			Voyager, Helios, and IMP 8 observations}}.
	\newblock \emph{\bibinfo{journal}{\jgr}} \textbf{\bibinfo{volume}{86}},
	\bibinfo{pages}{6673--6684} (\bibinfo{year}{1981}).
	
	\bibitem{Klein&Burlaga1982}
	\bibinfo{author}{{Klein}, L.~W.} \& \bibinfo{author}{{Burlaga}, L.~F.}
	\newblock \bibinfo{title}{{Interplanetary magnetic clouds at 1 AU}}.
	\newblock \emph{\bibinfo{journal}{\jgr}} \textbf{\bibinfo{volume}{87}},
	\bibinfo{pages}{613--624} (\bibinfo{year}{1982}).
	
	\bibitem{Marubashi2015}
	\bibinfo{author}{{Marubashi}, K.} \& \bibinfo{author}{{Cho}, K.~S.}
	\newblock \bibinfo{title}{{Non-Uniqueness of the Geometry of Interplanetary
			Magnetic Flux Ropes Obtained from Model-Fitting}}.
	\newblock \emph{\bibinfo{journal}{Sun and Geosphere}}
	\textbf{\bibinfo{volume}{10}}, \bibinfo{pages}{119--125}
	(\bibinfo{year}{2015}).
	
	\bibitem{Cho2017}
	\bibinfo{author}{{Cho}, K.~S.} \emph{et~al.}
	\newblock \bibinfo{title}{{Impact of the Icme-Earth Geometry on the Strength of
			the Associated Geomagnetic Storm: The September 2014 and March 2015 Events}}.
	\newblock \emph{\bibinfo{journal}{Journal of Korean Astronomical Society}}
	\textbf{\bibinfo{volume}{50}}, \bibinfo{pages}{29--39}
	(\bibinfo{year}{2017}).
	
	\bibitem{WangYM2015}
	\bibinfo{author}{{Wang}, Y.}, \bibinfo{author}{{Zhou}, Z.},
	\bibinfo{author}{{Shen}, C.}, \bibinfo{author}{{Liu}, R.} \&
	\bibinfo{author}{{Wang}, S.}
	\newblock \bibinfo{title}{{Investigating plasma motion of magnetic clouds at 1
			AU through a velocity-modified cylindrical force-free flux rope model}}.
	\newblock \emph{\bibinfo{journal}{Journal of Geophysical Research (Space
			Physics)}} \textbf{\bibinfo{volume}{120}}, \bibinfo{pages}{1543--1565}
	(\bibinfo{year}{2015}).
	
	\bibitem{HuQ2002}
	\bibinfo{author}{{Hu}, Q.} \& \bibinfo{author}{{Sonnerup}, B. U.~{\"O}.}
	\newblock \bibinfo{title}{{Reconstruction of magnetic clouds in the solar wind:
			Orientations and configurations}}.
	\newblock \emph{\bibinfo{journal}{Journal of Geophysical Research (Space
			Physics)}} \textbf{\bibinfo{volume}{107}}, \bibinfo{pages}{1142}
	(\bibinfo{year}{2002}).
	
	\bibitem{Zhang2013}
	\bibinfo{author}{{Zhang}, J.}, \bibinfo{author}{{Hess}, P.} \&
	\bibinfo{author}{{Poomvises}, W.}
	\newblock \bibinfo{title}{{A Comparative Study of Coronal Mass Ejections with
			and Without Magnetic Cloud Structure near the Earth: Are All Interplanetary
			CMEs Flux Ropes?}}
	\newblock \emph{\bibinfo{journal}{\solphys}} \textbf{\bibinfo{volume}{284}},
	\bibinfo{pages}{89--104} (\bibinfo{year}{2013}).
	
	\bibitem{Burlaga2002}
	\bibinfo{author}{{Burlaga}, L.~F.}, \bibinfo{author}{{Plunkett}, S.~P.} \&
	\bibinfo{author}{{St. Cyr}, O.~C.}
	\newblock \bibinfo{title}{{Successive CMEs and complex ejecta}}.
	\newblock \emph{\bibinfo{journal}{Journal of Geophysical Research (Space
			Physics)}} \textbf{\bibinfo{volume}{107}}, \bibinfo{pages}{1266}
	(\bibinfo{year}{2002}).
	
	\bibitem{Riley1997}
	\bibinfo{author}{{Riley}, P.}, \bibinfo{author}{{Gosling}, J.~T.} \&
	\bibinfo{author}{{Pizzo}, V.~J.}
	\newblock \bibinfo{title}{{A two-dimensional simulation of the radial and
			latitudinal evolution of a solar wind disturbance driven by a fast,
			high-pressure coronal mass ejection}}.
	\newblock \emph{\bibinfo{journal}{\jgr}} \textbf{\bibinfo{volume}{102}},
	\bibinfo{pages}{14677--14686} (\bibinfo{year}{1997}).
	
	\bibitem{Awasthi2018}
	\bibinfo{author}{{Awasthi}, A.~K.}, \bibinfo{author}{{Liu}, R.},
	\bibinfo{author}{{Wang}, H.}, \bibinfo{author}{{Wang}, Y.} \&
	\bibinfo{author}{{Shen}, C.}
	\newblock \bibinfo{title}{{Pre-eruptive Magnetic Reconnection within a
			Multi-flux-rope System in the Solar Corona}}.
	\newblock \emph{\bibinfo{journal}{\apj}} \textbf{\bibinfo{volume}{857}},
	\bibinfo{pages}{124} (\bibinfo{year}{2018}).
	
	\bibitem{Lynch2008}
	\bibinfo{author}{{Lynch}, B.~J.}, \bibinfo{author}{{Antiochos}, S.~K.},
	\bibinfo{author}{{DeVore}, C.~R.}, \bibinfo{author}{{Luhmann}, J.~G.} \&
	\bibinfo{author}{{Zurbuchen}, T.~H.}
	\newblock \bibinfo{title}{{Topological Evolution of a Fast Magnetic Breakout
			CME in Three Dimensions}}.
	\newblock \emph{\bibinfo{journal}{\apj}} \textbf{\bibinfo{volume}{683}},
	\bibinfo{pages}{1192--1206} (\bibinfo{year}{2008}).
	
\end{thebibliography}
\normalsize
\end{document}